\RequirePackage{fix-cm}
\documentclass[twocolumn,epjc3]{svjour3}  
\smartqed  
\usepackage[dvipdfmx]{graphicx}  
\RequirePackage[T1]{fontenc}
\RequirePackage{mathptmx}

\usepackage{bm}
\usepackage{float}
\usepackage{here}
\usepackage{enumerate}
\usepackage{makeidx}
\usepackage{extarrows}
\usepackage{fancybox}
\usepackage{fancyhdr}
\usepackage{indentfirst}
\usepackage{cite}
\usepackage{color}

\journalname{Eur. Phys. J. C}
\begin{document}

\title{A new relativistic viscous hydrodynamics code and its application to the Kelvin-Helmholtz instability in high-energy heavy-ion collisions
}


\author{Kazuhisa Okamoto \thanksref{addr1,e1}
        \and Chiho Nonaka \thanksref{addr1, addr2, addr3,e2}
}

\thankstext{e1}{e-mail: okamoto@hken.phys.nagoya-u.ac.jp}
\thankstext{e2}{e-mail: nonaka@hken.phys.nagoya-u.ac.jp}


\institute{Department of Physics, Nagoya University, Nagoya 464-8602, Japan \label{addr1}
           \and Kobayashi-Maskawa Institute for the Origin of Particles and the Universe (KMI), Nagoya University, Nagoya 464-8602, Japan \label{addr2} 
           \and Department of Physics, Duke University, Durham, NC, 27708, USA \label{addr3}     
}
\date{Received: date / Accepted: date}
\maketitle
\begin{abstract}
We construct a new relativistic viscous hydrodynamics code optimized in the Milne coordinates. 
We split the conservation equations into an ideal part and a viscous part, using the Strang spitting method. 
In the code a Riemann solver based on the two-shock approximation is utilized for the ideal part 
and the Piecewise Exact Solution (PES)  method is applied for the viscous part. 
We check the validity of our numerical calculations by comparing analytical solutions, the 
viscous Bjorken's flow and the Israel-Stewart theory in Gubser flow regime. 
Using the code, we discuss possible development of the Kelvin-Helmholtz instability 
in high-energy heavy-ion collisions. 
\end{abstract}

\keywords{Relativistic heavy-ion collisions \and Relativistic hydrodynamics \and Numerical hydrodynamics \and Kelvin-Helmholtz instability}
\PACS{25.75.-q \and 47.75.+f \and 47.11.-j \and 47.20.Ft }

\section{Introduction}

Since the success of production of the strongly interacting quark-gluon plasma (QGP) at 
Relativistic Heavy Ion Collider (RHIC) \cite{QGP-RHIC},  
relativistic viscous hydrodynamic model has been one of promising phenomenological models. 
Now at RHIC as well as at the Large Hadron Collider (LHC) high-energy heavy-ion collisions are 
carried out. The strong collective dynamics observed in experimental data at RHIC and the LHC provides us with 
a clue of understanding the QCD matter. 
A relativistic hydrodynamic model  is suitable for description of space-time evolution of 
strongly interacting QCD matter produced after collisions. 
Besides, it  has a close relation to an equation of state and transport coefficients of the QGP. 
The QCD phase transition mechanism and the QGP bulk property is elucidated 
from comparison between hydrodynamic calculation and experimental data. 

A  relativistic viscous hydrodynamic model plays an important role in the quantitative understanding of the QGP bulk property. 
However, introducing viscosity effect into the framework of relativistic hydrodynamics is not an easy task, because of   
the acausality problem. 
There is not the unique way to extract the second-order relativistic viscous hydrodynamic equation. 
In high-energy heavy-ion collisions, currently the Israel-Stewart theory \cite{Stewart1977,Israel1979} and conformal 
hydrodynamics  \cite{Baier2007} are often used. 
Solving them numerically, study of experimental data of high-energy heavy-ion collisions 
is performed \cite{Song2007, Dusling2007, Luzum2008, Denicol2009, Schenke2011,Bozek2011,Roy2012,Karpenko2014,Noronha2014,Pang2014}. 

Now the relativistic viscous hydrodynamic model can explain not only the elliptic flow 
but also higher harmonics \cite{Gale2013}. 
In particular,  analyses of the higher harmonics bring us progress of understanding of  the QGP,  
because it is more sensitive to the QGP bulk property. 
Furthermore, a lot of experimental data are reported; correlation between flow harmonics \cite{ATLAS2015, ALICE2016}, 
event plane correlation \cite{ATLAS2014, CMS2015}, non-linearity of higher flow harmonics  \cite{Yan2015} and  
three particle correlation \cite{STAR2017a, STAR2017b}. 
At the same time, we can investigate the QGP property further using information of (3+1)-dimensional space-time 
evolutions \cite{CMS2015, STAR2017a, STAR2017b}. 
The rich experimental data realizes investigation of both shear and bulk viscosities and even their temperature dependence. 

We  need to perform numerical calculations for relativistic viscous hydrodynamics with high accuracy,
 to achieve the quantitative analyses of the transport coefficients of the QGP from comparison with high statistics and high precision 
experimental data. 
For example, the following features in numerical calculations are demanded:  A fluctuating initial condition is correctly captured 
and numerical viscosity which is needed for stability of calculation is much smaller than physical viscosity. 
Furthermore, time evolution of the viscous stress tensor is sensitive to numerical scheme, because 
it consists of time and space derivatives of hydrodynamic variables. 

Here we  present a new relativistic viscous hydrodynamics code optimized in the Milne coordinates.  
The code is developed based on our algorithm of the ideal fluid in which a Riemann solver with the two shock 
approximation \cite{Mignone2005} is employed \cite{Okamoto2016}. 
It is stable even with small numerical viscosity \cite{Akamatsu2014}. 
We shall show comparison between numerical calculations and analytic solutions of viscous Bjorken's flow 
and the Israel-Stewart theory in Gubser flow regime. 

Using the code, we shall discuss possible development of Kelvin-Helmholtz (KH)  instability in high-energy heavy-ion collisions. 
Hydrodynamic instability and turbulent flow are discussed in Ref.\:\cite{Romatschke2008, Floerchinger2011} and 
the possibility of KH instability is argued in Ref.\:\cite{Csernai2012}. 
The hydrodynamic instability is affected by a viscosity effect, which suggests that the numerical code with less numerical 
viscosity is indispensable for study of it. 

This paper is organized as follows. We begin in Sect.\:\ref{Sec:hydro}  by showing the relativistic viscous hydrodynamic equations briefly.  
In Sect.\:\ref{Sec:algorithm}  we explain the numerical algorithm; Strang splitting method and numerical implementation. 
We check the validity of our code comparing analytic solutions of viscous Bjorken flow and the Israel-Stewart theory in the Gubser flow 
regime in Sect.\:\ref{Sec:test}. 
In Sect.\:\ref{Sec:KHI}, we discuss the possible development of KH instability in high-energy heavy-ion collisions. 
We end in Sect.\:\ref{Sec:sum} with our conclusions.

\section{Relativistic viscous hydrodynamic equations \label{Sec:hydro}}

The relativistic hydrodynamics is based on the conservation equations,  
\begin{align} N^\mu_{\; ;\mu}  &= 0 , \label{eq:cons1} \\
T^{\mu\nu}_{\;\;\; ;\mu} &=0 \label{eq:cons2}, 
\end{align}
where $N^\mu$ is the net charge current and $T^{\mu\nu}$ is the energy-momentum tensor. 
In the case of ideal fluid, the net charge current and energy-momentum tensor  are given by
\begin{align} N^\mu & = n u^\mu , \label{eq:id-tensor1} \\
T^{\mu\nu} &= e u^\mu u^\nu - p \Delta^{\mu\nu} ,   \label{eq:id-tensor2} 
\end{align}
where $n$ is the net charge density, $e$ is the energy density, $p$ is the pressure and $u^\mu$ is the fluid four-velocity 
which satisfies the normalization $u^\mu u_\mu = 1$. 
$\Delta^{\mu\nu}$ is the orthogonal projection tensor to $u^\mu$, which is 
defined by 
\begin{align} \Delta^{\mu\nu} = g^{\mu\nu} - u^\mu u^\nu ,
\end{align}
with the metric tensor $g^{\mu\nu}$. 
Here the $u^\mu$ is determined uniquely. 

On the other hand, in dissipative flow, there are several possible choices to determine $u^\mu$. 
For example, one can assign the $u^\mu$ as net charge flow (Eckart frame \cite{Eckart1940}) 
or as energy flow (Landau frame \cite{Landau1987}).
The decomposition of $N^\mu$ and $T^{\mu\nu}$  in viscous fluid depends on the choice of $u^\mu$. 
Here we choose the Landau frame for relativistic viscous hydrodynamic equations, because we focus on the 
high-energy heavy-ion collisions as RHIC and the LHC where the net baryon number is very small \cite{Andronic2013}.

In the Landau frame, the net charge current and the energy-momentum tensor of the viscous fluid are decomposed as 
\begin{align}
N^\mu & = n u^\mu + n^\mu, \label{eq:tensor1}\\
T^{\mu\nu}& =  eu^\mu u^\nu - (p+ \Pi) \Delta^{\mu\nu} + \pi^{\mu\nu},   \label{eq:tensor2} 
\end{align}
where $n^\mu$ is the charge diffusion current, $\Pi$ is the bulk pressure, and $\pi^{\mu\nu}$ is the shear tensor \cite{Landau1987}.   
The relativistic extension of Navier-Stokes theory in non-relativistic fluid  usually has a problem of acausality and instability
 \cite{Hiscock1983,Hiscock1985,Pu2010}. 
The problem can be resolved by introducing the second-order terms of the viscous tensor and the derivative of fluid variables 
into the hydrodynamic equations \cite{Stewart1977,Israel1979}. 
However, the original Israel-Stewart theory does not reproduce the results 
of the kinetic equation quantitatively \cite{Huovinen2009,Bouras2010, Molnar2009,Takamoto2010,Xu2010}.  
The construction of second-order relativistic viscous hydrodynamic equations is still 
under investigation. 
The extension of the Israel-Stewart theory is also  proposed 
\cite{Muronga2007,Betz2009,Denicol2010,Calzetta2010, Denicol2012, Jaiswal2013}. 
In addition to the framework of the Israel-Stewart theory, other approaches such as the AdS/CFT correspondence \cite{Baier2007, Natsuume2008,Bhatta2008, 
Romatschke2010} and renormalization group method are applied to the construction of causal 
relativistic hydrodynamics \cite{Tsumura2012,Tsumura2015}.

In the second-order viscous hydrodynamics, additional equations for evolution of the viscous 
tensors are needed. 
Here, we introduce the convective time derivative $D$ and the spatial gradient operator $\nabla^\mu$, which are defined by 
\begin{align} &DA^{\mu_1 \cdots\mu_n} \equiv u^\beta A^{\mu_1\cdots\mu_n}_{; \beta} , \\
& \nabla_\alpha A^{\mu_1 \cdots\mu_n} \equiv \Delta^\beta_\alpha A^{\mu_1\cdots\mu_n}_{; \beta}, 
\end{align}
respectively. 
For example, in the second-order Israel-Stewart formalism 
the constitutive equations of the viscous tensors are given by
\begin{align} \Delta^{\mu}_{\;\;\alpha} D n^\alpha &= -\frac{1}{\tau_n} (n^\mu - n^\mu_{\rm NS}) - I_n^\mu , \label{eq:IS1} \\
\Delta^\mu_{\;\;\alpha}\Delta^\nu_{\;\;\beta}D\pi^{\alpha\beta} &= -\frac{1}{\tau_\pi} (\pi^{\mu\nu} - \pi^{\mu\nu}_{\rm NS}) - I^{\mu\nu}_{\pi} ,   \label{eq:IS2} \\
D\Pi & = -\frac{1}{\tau_\Pi} (\Pi - \Pi_{\rm NS}) - I_\Pi , \label{eq:IS3}
\end{align}
where $\tau_n, \tau_\pi$, and $\tau_\Pi$ are relaxation times, $I^\mu_n, I^{\mu\nu}_\pi,$ and $I_\Pi$ 
represent second-order terms. $n^\mu_{\rm NS}, \pi^{\mu\nu}_{\rm NS}$, and $\Pi_{\rm NS}$ are the Navier-Stokes value of viscous tensors written as 
\begin{align} n_{\rm NS}^\mu &= \sigma T\nabla_\mu \left(\frac{\mu}{T}\right),  \\
\pi^{\mu\nu}_{\rm NS}  &=  \eta \left( \nabla^\mu u^\nu + \nabla^\nu u^\mu - \frac{2}{3}\Delta^{\mu\nu}\theta  \right) , \\ 
\Pi_{NS} &=-\zeta\theta, 
\end{align}
where $T$ is the temperature, $\mu$ is the chemical potential, $\theta\equiv u^\mu_{;\mu}$ is the expansion scalar, $
\sigma$ is the charge conductivity, $\eta$ is the shear viscosity, and $\Pi$ is the bulk viscosity. 

\par We construct a relativistic viscous hydrodynamics code in the Milne coordinates 
$(\tau,x,y,\eta)$ which is  optimized for description of the strong longitudinal expansion \cite{Bjorken1983}  at RHIC and the LHC. 
In the Milne coordinates, the metric tensor is given by $g^{\mu\nu}={\rm diag}(1, -1,-1,-1/\tau^2)$ and 
the fluid four-velocity has the form $u^\mu = \gamma(1, v^x, v^y,v^\eta)$, where $v^i (i=x,y,\eta)$ 
and $\gamma= (1- {v^x}^2 - {v^y}^2 - \tau^2 {v^\eta}^2)^{-1/2}$  are the three-velocity and the Lorentz factor, respectively. 
The conservation equations Eqs.\:\eqref{eq:cons1} and \eqref{eq:cons2}  are explicitly written as
\begin{align} & \partial_\tau N^\tau +  \partial_i N^i = -\frac{1}{\tau} N^\tau, \label{eq:consMilne1} \\
& \partial_\tau T^{\tau \nu} + \partial_i T^{i\nu} = S^\nu ,\label{eq:consMilne2}
\end{align}
where $i=x,y, \eta$ and the right-hand sides of them represent geometric source terms. 
$S^\nu$ is given by 
\begin{equation} 
S^\nu =\left( -\frac{1}{\tau}T^{\tau\tau} -\tau T^{\eta\eta},\: - \frac{1 }{\tau}T^{\tau x},\: - \frac{1}{\tau}T^{\tau y}, \:-\frac{3}{\tau}T^{\tau\eta} \right). 
\end{equation}
The constitutive equations Eqs.~\eqref{eq:IS1},  \eqref{eq:IS2},  and \eqref{eq:IS3} in Milne coordinates read 
\begin{align} (\partial_\tau + v^i\partial_i)n^\mu & = -\frac{1}{\gamma\tau_n} (n^\mu - n^\mu_{\rm NS}) - I_n^\mu - J^\mu_n - K^\mu_n , \label{eq:ISMilne1} \\
 (\partial_\tau + v^i\partial_i)\pi^{\mu\nu} & = - \frac{1}{\gamma  \tau_{\eta}} (\pi^{\mu\nu} - \pi^{\mu\nu}_{NS}) 
 - I^{\mu\nu}_{\pi} -J^{\mu\nu}_\pi - K^{\mu\nu}_\pi , \label{eq:ISMilne2} \\
(\partial_\tau + v^i\partial_i)\Pi & = -\frac{1}{\gamma\tau_\Pi} (\Pi - \Pi_{\rm NS}) - I_\Pi , \label{eq:ISMilne3} 
\end{align}
where $\tau_n$, $\tau_\eta$ and $\tau_\Pi$ are the relaxation times, and the second-order terms are defined by
\begin{align}
& J_n^\tau = \tau v^\eta n^\eta,\quad  J_n^j = 0 , \\
& J_n^\eta = \frac{1}{\tau}v^\eta n^\tau + \frac{1}{\tau} n^\eta , \\
& J^{\tau\tau}_\pi = 2\tau v^\eta\pi^{\tau\eta} , 
 \quad J^{\tau j}_\pi =  \tau v^\eta \pi^{j \eta}, \\
& J_\pi^{xx} = J_\pi^{yy} = J_\pi^{xy} = 0 , \\
& J_\pi^{i\eta} =  \frac{1}{\tau}\pi^{i\eta} + \frac{1}{\tau} v^\eta \pi^{i\tau}  , \\
& J_\pi^{\eta\eta} = \frac{2}{\tau}v^\eta\pi^{\tau\eta} + \frac{2}{\tau} \pi^{\eta\eta}, \\
& K^{\mu}_n = n^\lambda v^\mu Du_\lambda , \\
&K^{\mu\nu}_\pi = (\pi^{\lambda \mu} v^\nu + \pi^{\lambda\nu}v^\mu)Du_\lambda,
\end{align}
$j = x, y$ and $\lambda = \tau,x,y,\eta$. 
Here, $J^\tau_n$ and $J^{\mu\nu}_\pi$ are the geometric source terms which come from the convective time derivative of $n^\mu$ and $\pi^{\mu\nu}$ respectively.
$K^\mu_n$ and $K^{\mu\nu}_\pi$ ensures the constraints $n^\mu u_\mu=0, \pi^{\mu\nu}u_\mu=0$ and $\pi^\mu_\mu =0$.

\section{Numerical algorithm \label{Sec:algorithm}}
In this section, we present our numerical algorithm for solving the relativistic viscous 
hydrodynamic equations in the Milne coordinates. 

\subsection{Strang splitting method}
In our algorithm, we split the conservation equations Eqs.\:\eqref{eq:consMilne1} and \eqref{eq:consMilne2} 
into two parts, an ideal part and a viscous part using the Strang splitting method \cite{Strang1968}. 
Specifically, the net charge current and the energy-momentum tensor are divided 
as follows: $N^\mu= N^\mu_{\rm id} + N^\mu_{\rm vis}$ and $T^{\mu\nu} = T^{\mu\nu}_{\rm id} + T^{\mu\nu}_{\rm vis}$, 
where $N^\mu_{\rm id} \equiv n u^\mu, N^\mu_{\rm vis}\equiv n^\mu$,  
$T^{\mu\nu}_{\rm id} \equiv eu^\mu u^\nu - p \Delta^{\mu\nu}$ and $T^{\mu\nu}_{\rm vis} \equiv \pi^{\mu\nu} - \Pi \Delta^{\mu\nu}$. 
The subscripts ``id'' and ``vis'' mean the  ideal part and the viscous part, respectively.  
The equations of the ideal part are expressed by
 \begin{align}
& \partial_\tau N^\tau_{\rm id} +  \partial_i N^i_{\rm id} = -\frac{1}{\tau} N^\tau_{\rm id},  \label{eq:id1}\\
& \partial_\tau T^{\tau \nu}_{\rm id} + \partial_i T^{i\nu}_{\rm id} = S^\nu_{\rm id},  \label{eq:id2} 
\end{align}
where $S^\nu_{\rm id} =\left( -T^{\tau\tau}_{\rm id}/\tau -\tau T^{\eta\eta}_{\rm id} ,\: - T^{\tau x}_{\rm id}/\tau, \: - T^{\tau y}_{\rm id}/\tau, \:-3T^{\tau\eta}_{\rm id}/\tau \right)$. 
They are nothing but usual ideal hydrodynamic equations in the Milne coordinates.  
On the other hand, the equations of the viscous part  are given by 
\begin{align}
& \partial_\tau (N^\tau_{\rm id}+N^\tau_{\rm vis} )+  \partial_i N^i_{\rm vis} = -\frac{1}{\tau} N^\tau_{\rm vis} , \label{eq:vis1} \\
& \partial_\tau(T^{\tau \nu}_{\rm id} + T^{\tau \nu}_{\rm vis}) + \partial_i T^{i\nu}_{\rm vis} = S^\nu_{\rm vis} ,  \label{eq:vis2} 
\end{align}
where $S^\nu_{\rm vis} =\left( -T^{\tau\tau}_{\rm vis}/\tau - \tau T^{\eta\eta}_{\rm vis} , 
\: - T^{\tau x}_{\rm vis}/\tau, \: - T^{\tau y}_{\rm vis}/\tau, \:-3T^{\tau\eta}_{\rm vis}/\tau \right)$. 
They give viscous corrections to the evolution of the ideal fluid. 

\par The Strang splitting technique is also applied to evaluate the constitutive equations of the viscous tensors 
Eqs.\:\eqref{eq:ISMilne1}-\eqref{eq:ISMilne3}.
We decompose the constitutive equations into the following three parts; 
the convection equations, 
\begin{align} (\partial_\tau + v^i\partial_i)n^\mu & = 0 , \label{eq:conv1} \\
 (\partial_\tau + v^i\partial_i)\pi^{\mu\nu} & =0 , \label{eq:conv2} \\
(\partial_\tau + v^i\partial_i)\Pi & = 0, \label{eq:conv3} 
\end{align}
the relaxation equations, 
\begin{align} \partial_\tau n^\mu & = -\frac{1}{\gamma\tau_n} (n^\mu - n^\mu_{\rm NS})  , \label{eq:relax1} \\
 \partial_\tau \pi^{\mu\nu} & = - \frac{1}{\gamma  \tau_{\eta}} (\pi^{\mu\nu} - \pi^{\mu\nu}_{NS}) 
 , \label{eq:relax2} \\
\partial_\tau \Pi & = -\frac{1}{\gamma\tau_\Pi} (\Pi - \Pi_{\rm NS}), \label{eq:relax3} 
\end{align}
and the equations with source terms, 
\begin{align} \partial_\tau n^\mu & = - I_n^\mu - J^\mu_n - K^\mu_n , \label{eq:source1} \\
 \partial_\tau \pi^{\mu\nu} & = - I^{\mu\nu}_{\pi} -J^{\mu\nu}_\pi - K^{\mu\nu}_\pi , \label{eq:source2} \\
\partial_\tau \Pi & = - I_\Pi.  \label{eq:source3} 
\end{align}
In numerical simulation of relativistic hydrodynamic equation, a time-step size $\Delta \tau$ is usually 
determined by the Courant-Friedrichs-Lewy (CFL) condition. 
However, in the relativistic dissipative hydrodynamics,   
one needs to determine the value of $\Delta \tau$ carefully.  
The relaxation times $\tau_n, \tau_\eta$, and $\tau_\Pi$ in the constitutive  equations show 
the characteristic timescale of  evolutions of the viscous tensors, which means that 
a small relaxation time gives us more restrictive condition to $\Delta \tau$ than the CFL condition does. 
If the relaxation times $\tau_n, \tau_\eta$, and $\tau_\Pi$ are much shorter than the fluid timescale $\tau_{\rm fluid}$,  
the time-step size $\Delta\tau$ should be smaller than the relaxation timescale, 
which makes the computational cost increase. 
To avoid this problem, we use the Piecewise Exact Solution (PES) method  \cite{Inutsuka2011}, instead of using a simple explicit scheme. 
In the PES method, formal solutions of Eqs.\:\eqref{eq:relax1}-\eqref{eq:relax3}, 
\begin{align}n^\mu(\tau) &= (n^\mu_0 - n^\mu_{\rm NS}){\rm exp} \left[ - \frac{\tau - \tau_0}{\gamma\tau_n}\right] + n^\mu_{\rm NS} , \\
\pi^{\mu\nu}(\tau) &= (\pi^{\mu\nu}_0 - \pi^{\mu\nu}_{\rm NS}){\rm exp} \left[ - \frac{\tau - \tau_0}{\gamma\tau_\pi}\right] + \pi^{\mu\nu}_{\rm NS} ,\\
 \Pi(\tau) &= (\Pi_0 - \Pi_{\rm NS}){\rm exp} \left[ - \frac{\tau - \tau_0}{\gamma\tau_\Pi}\right] +\Pi_{\rm NS}, 
\end{align}
can be used. 
On the other hands, if the relaxation times are larger than $\Delta\tau$ determined by the CFL condition, 
the PES method is not applied \cite{Karpenko2014}. 

In our algorithm, we solve the time evolution of $n^x, n^y, n^\eta, \pi^{xx}, \pi^{yy}, \pi^{\eta\eta},\pi^{xy},\pi^{y\eta}, \pi^{\eta x}$ and $\Pi$ directly. 
Other components of viscous tensors  
$n^\tau, \pi^{\tau\tau}, \pi^{\tau x}, \pi^{\tau y}$ and $\pi^{\tau \eta}$ are derived from the orthogonality conditions 
$n^\mu u_\mu=0$ and $\pi^{\mu\nu} u_{\nu}=0$.

\subsection{Numerical implementation}
The decomposed hydrodynamic equations Eqs.\:\eqref{eq:id1}-\eqref{eq:source3} are solved by 
the following procedure. 
Here, we represent a conserved variable as $\bm U = \bm U_{\rm id} + \bm U_{\rm vis}$, 
where $\bm U_{\rm id}\equiv (N^\tau_{\rm id}, T^{\tau\nu}_{\rm id})$ and 
$\bm U_{\rm vis} \equiv (N^\tau_{\rm vis}, T^{\tau\nu}_{\rm vis})(\nu=\tau,x,y,\eta)$. 
Fluid and dissipative variables are described by $\bm V_{\rm id}\equiv (n, p,v^i)$ 
and  $\bm V_{\rm vis}\equiv(n^i, \pi^{ij}, \Pi) (i,j=x,y,\eta)$, respectively. 

\par First, we solve the ideal part of the conservation equations Eqs.\:\eqref{eq:id1} and \eqref{eq:id2} 
using the Riemann solver \cite{Okamoto2016}.
In this step, the conserved 
variable $\bm U_{\rm id}(\tau)$ is  evolved into 
$\bm U_{\rm id}^*(\tau+\Delta\tau)$, 
where the asterisk indicates a variable evolved only in the ideal part. 
$\bm V_{\rm id}(\tau)$ is used to evaluate the numerical flux and the geometric source terms 
in Eqs.\:\eqref{eq:id1} and \eqref{eq:id2}. 
We calculate the fluid variable $\bm V_{\rm id}^*(\tau+\Delta\tau)$ from $\bm U^*_{\rm id}(\tau+\Delta\tau)$ with 
the algorithm for recovery of the primitive variables $\bm V_{\rm id}$ from the conserved variables 
$\bm U_{\rm id}$ \cite{Akamatsu2014}. 

\par Second, we solve the constitutive equations of the viscous tensors Eqs.\:\eqref{eq:conv1}-\eqref{eq:source3} to 
obtain $\bm V_{\rm vis}(\tau+\Delta\tau)$. 
The convection equations Eqs.\:\eqref{eq:conv1}-\eqref{eq:conv3}, 
the relaxation equations Eqs.\:\eqref{eq:relax1}-\eqref{eq:relax3} and Eqs.\:\eqref{eq:source1}-\eqref{eq:source3} 
are solved by the upwind scheme, the PES method and the predictor corrector method, respectively. 
The Navier-Stokes terms $n^\mu_{\rm NS}, \pi^{\mu\nu}_{\rm NS}, \Pi_{\rm NS}$ and the second-order terms $I$, $K$ in the right-hand sides of Eqs.\:\eqref{eq:relax1}-\eqref{eq:source3} contain not only the spatial derivatives of fluid variables but also the time derivatives of them. 
The time derivatives in the right-hand sides of Eqs.\:\eqref{eq:relax1}-\eqref{eq:source3} are obtained by $\partial_\tau \bm V_{\rm id} = (\bm V^*_{\rm id}(\tau +\Delta\tau) - \bm V_{\rm id}(\tau) )/ \Delta\tau.$
Here we keep the middle time-step value of the viscous tensor $\bm V_{\rm vis}(\tau+\Delta\tau/2)$ for the next step.

\par Next, the conserved variables $\bm U^*_{\rm id}(\tau+\Delta\tau)$ and $\bm U_{\rm vis}(\tau)$ are 
evolved into  $\bm U_{\rm id}(\tau+\Delta\tau)$  and  $\bm U_{\rm vis}(\tau+\Delta\tau)$  
by the viscous part of conservation equation Eqs.\:\eqref{eq:vis1} and \eqref{eq:vis2}. 
Then we recover the fluid variables $\bm V_{\rm id}(\tau + \Delta\tau)$ from conserved variables $\bm U_{\rm vis}(\tau+\Delta\tau)$  \cite{Inutsuka2011}.
We keep  the middle time-step value $\bm V_{\rm id}(\tau+\Delta/2)$.

\par To achieve the second-order accurate in time, we repeat the above whole steps 
using the middle time-step values $\bm V_{\rm id}(\tau+\Delta\tau/2)$ 
and $\bm V_{\rm vis}(\tau+\Delta\tau/2)$. 
However, we find that numerical errors arise mainly from the constitutive equations Eqs.\:\eqref{eq:conv1}-\eqref{eq:source3}.  
Therefore we carry out numerical calculation in the second-order accurate in time only in constitutive equations Eqs.\:\eqref{eq:conv1}-\eqref{eq:source3} and the viscous part of 
conservation equations Eqs.\:\eqref{eq:vis1} and \eqref{eq:vis2}. 

Throughout all above steps, we evaluate space derivative terms using the MC limiter \cite{Van1979} for the second-order accurate in space or 
the piecewise parabolic method (PPM)  \cite{Colella1984,Marti1996,Colella2008} for the third-order accurate in space. 
We shall give the explicit expressions of the interpolation procedures, the MC limiter and the PPM in \ref{app-interpolation}. 

\section{Numerical tests \label{Sec:test}}

We  check the correctness of our code in the following test problems;  
the viscous Bjorken flow for one-dimensional expansion and 
the Israel-Stewart theory in Gubser flow regime \cite{Marrochio2015} for  
the three-dimensional calculation.
We use the ideal massless gas equation of state, $p=e/3$ and set the net charge to be vanishing. 

\subsection{Viscous Bjorken flow}

The Bjorken flow is one of the simplest one-dimensional test problems for the code which is optimized in the Milne coordinates.
In the ideal fluid, the time evolution of temperature follows  $T=T_0(\tau_0/\tau)^{1/3}$, 
where $\tau_0$ and  $T_0$  are the initial proper time and  temperature, respectively \cite{Bjorken1983}. 
In the viscous fluid, 
the non-vanishing components of viscous tensor $\pi^{\mu\nu}$ 
are $\pi^{xx}$, $\pi^{yy}$, and $\pi^{\eta\eta}$. 
From the symmetries of the system, the relation $2\pi^{xx}=2\pi^{yy} = - \tau^2 \pi^{\eta\eta}$ holds. 
First, we focus on the shear viscosity effects at the Navier-Stokes limit.  
In the Navier-Stokes limit, the relativistic viscous hydrodynamic equation with the boost invariance is written as 
\begin{align} \frac{\partial e}{\partial \tau} = - \frac{e+p+\tau^2\pi^{\eta\eta}_{\rm NS}}{\tau},  \label{eq:eqBj}
\end{align}
where $\pi^{\eta\eta}_{\rm NS}$ is the Navier-Stokes value of shear tensor,  
\begin{align} \pi^{\eta\eta}_{\rm NS} = -\frac{4\eta}{3\tau^3}.  \label{eq:piBj}
\end{align}
If $\eta/s$ is constant, Eqs.\:\eqref{eq:eqBj} and \eqref{eq:piBj}  give the time evolution of the temperature, 
\begin{align} T = \left(\frac{\tau_0}{\tau} \right)^{1/3} \left[ T_0 + \frac{2}{3\tau_0}\frac{\eta}{s} \left( 1 - \left(\frac{\tau_0}{\tau}\right)^{2/3} \right) \right] .
\end{align}

\begin{figure}[t]
    \centering
  \includegraphics[width=8cm]{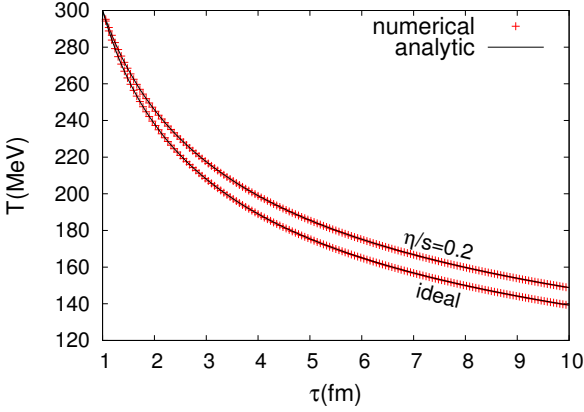}
   \centering
 \caption{The numerical and analytical  results of the time evolution of the temperature 
 in the Bjorken flow with and without shear viscosity. 
 The viscosity to entropy density ratio is $\eta/s=0$ and $0.2$. \label{fig:Bj-eta}} 
\end{figure}

The numerical calculation is carried out on the space-grid size $\Delta\eta = 0.1$ with the time-step size $\Delta\tau = 0.1 \tau_0\Delta\eta$. 
The initial temperature $T_0$ and the proper time $\tau_0$ are set to $T_0=300$ MeV and  $\tau_0 = 1$ fm, respectively. 
We set the relaxation time to be  $\tau_\eta= 0.0001$ fm as  the Navier-Stokes limit.  
Since $\tau_\eta$ is smaller than $\Delta\tau$, 
the PES method is applied to solve the relaxation equations Eqs.\:\eqref{eq:relax1}-\eqref{eq:relax3}. 
Figure \ref{fig:Bj-eta} shows the analytical and numerical results of the Bjorken flow with and without 
shear viscosity. 
In the case of finite shear viscosity, the temperature decreases with proper time more slowly, 
compared to that of the ideal fluid. 
In both cases, our numerical results show good agreement with the analytical solutions.

\begin{figure}[t!]
    \centering
  \includegraphics[width=8cm]{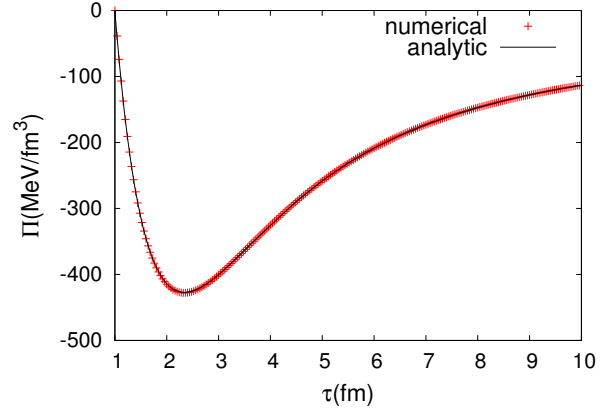}
   \centering
 \caption{The numerical and analytical results of the time evolution of the Bulk 
 pressure in the Bjorken flow. The bulk viscosity is $\zeta=1000$MeV/fm$^2$. \label{fig:Bj-bv}} 
\end{figure}

Next, 
we check  the time evolution of the bulk pressure in the viscous Bjorken's flow.  
Ignoring the second-order terms $I_\Pi$ in Eq.\: \eqref{eq:ISMilne3}, 
we write the relaxation equation of the bulk pressure,  
\begin{align} \frac{\partial\Pi}{\partial\tau} = -\frac{1}{\tau_\Pi} \left(\Pi - \Pi_{\rm NS}\right), \label{eq:relax-Bj}
\end{align}
with the Navier-Stokes value of the bulk pressure  $\Pi_{\rm NS} = \zeta/\tau$. 
If we assume $\zeta$ and $\tau_\Pi$ are constant, 
we obtain the analytical solution of  Eq.\:\eqref{eq:relax-Bj}, 
\begin{align} \Pi = \Pi_0 e^{-(\tau - \tau_0)/\tau_\Pi} + \frac{\zeta}{\tau_\Pi} e^{-\tau/\tau_\Pi}
\left[ {\rm Ei}(\tau_0/\tau_\Pi) - {\rm Ei}(\tau/\tau_\Pi)\right],
\end{align}
where $\Pi_0$ is the initial value of the bulk pressure and ${\rm Ei}(x)$ is the exponential integral function.

\begin{figure*}[ht]
  \begin{minipage}{0.5\hsize}
  \centering
  \includegraphics[width=8cm]{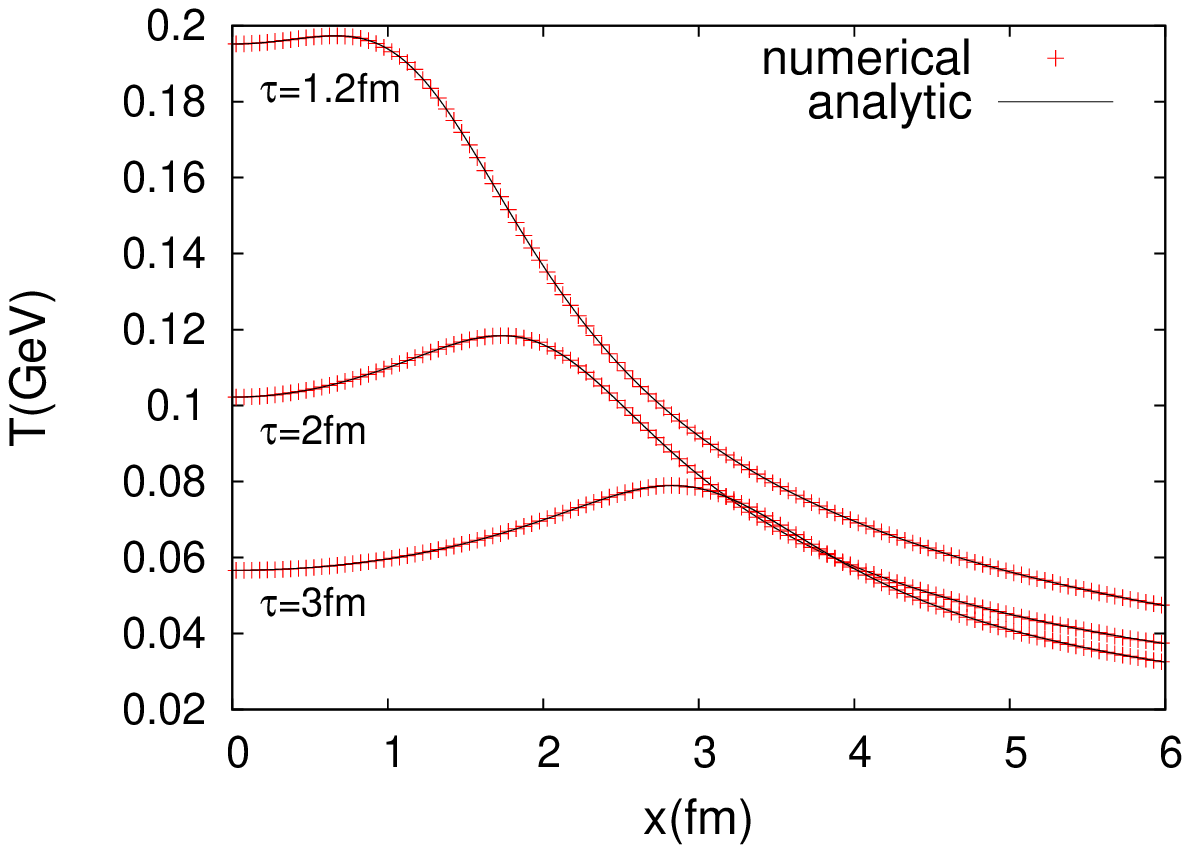}
  \end{minipage}
  \begin{minipage}{0.49\hsize}
  \centering
  \includegraphics[width=8cm]{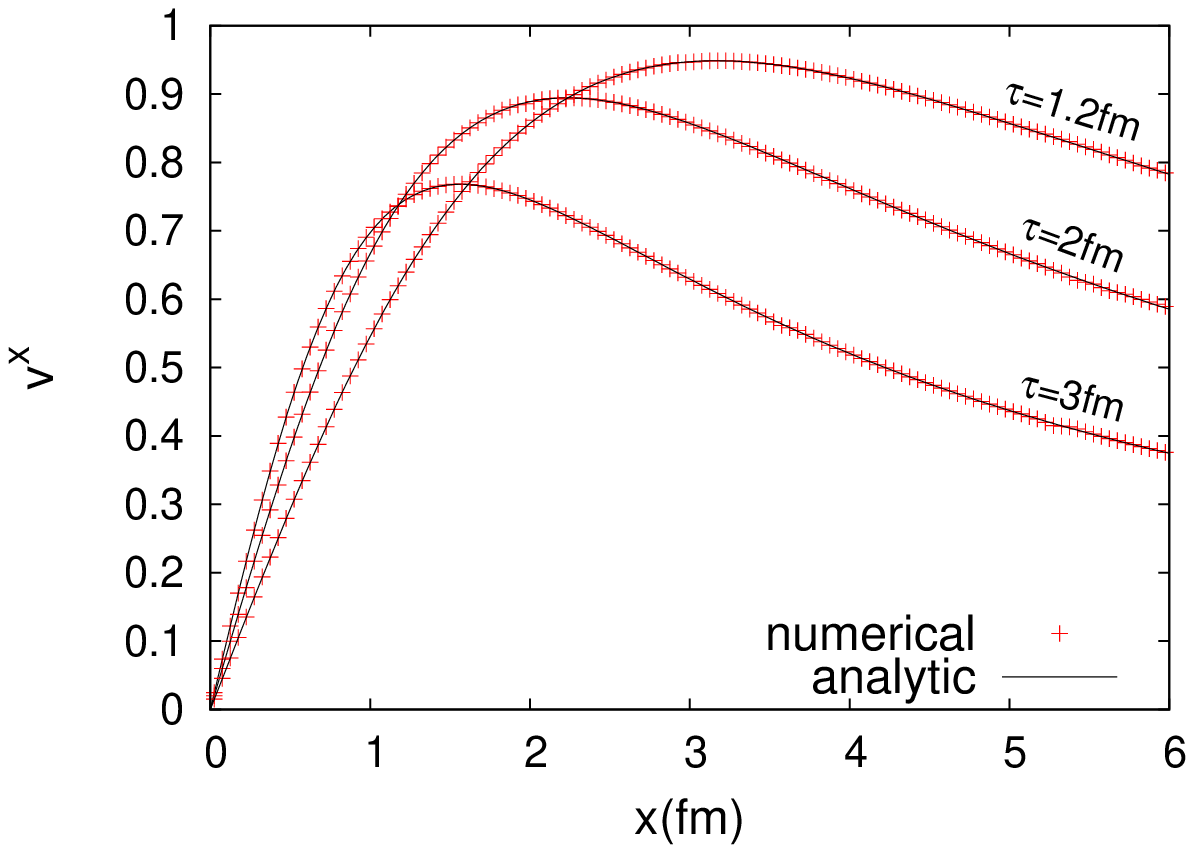}
  \end{minipage} 
  \caption{Comparison between the solutions for temperature $T$ (left panel) and the $x$ component of fluid velocity $v^x$ (right panel) 
  from  the Gubser flow and our numerical calculation as a function of $x$.
The solid lines stand for the semi-analytic solutions and  the pluses stand for numerical results. 
\label{fig:Gubser-T} }
\end{figure*}

\begin{figure*}[ht]
  \begin{minipage}{0.5\hsize}
  \centering
  \includegraphics[width=8cm]{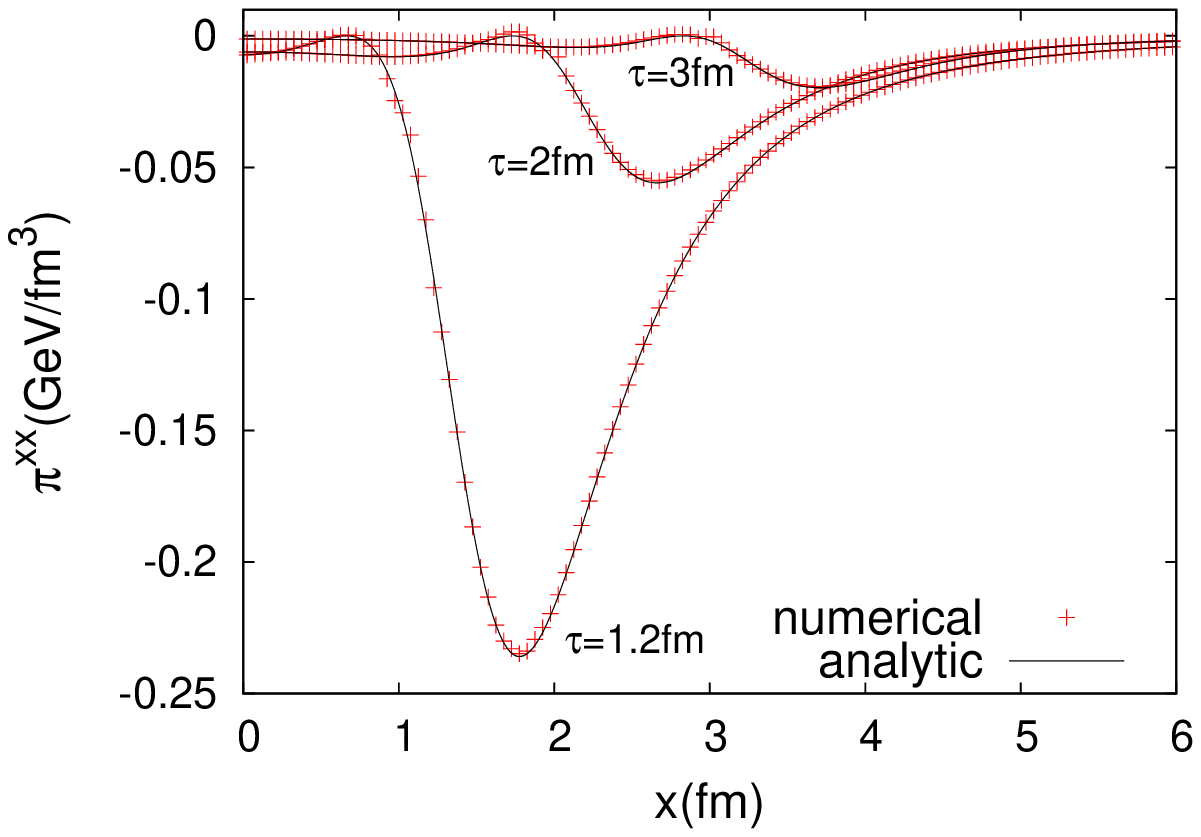}
  \end{minipage}
  \begin{minipage}{0.49\hsize}
  \centering
  \includegraphics[width=8cm]{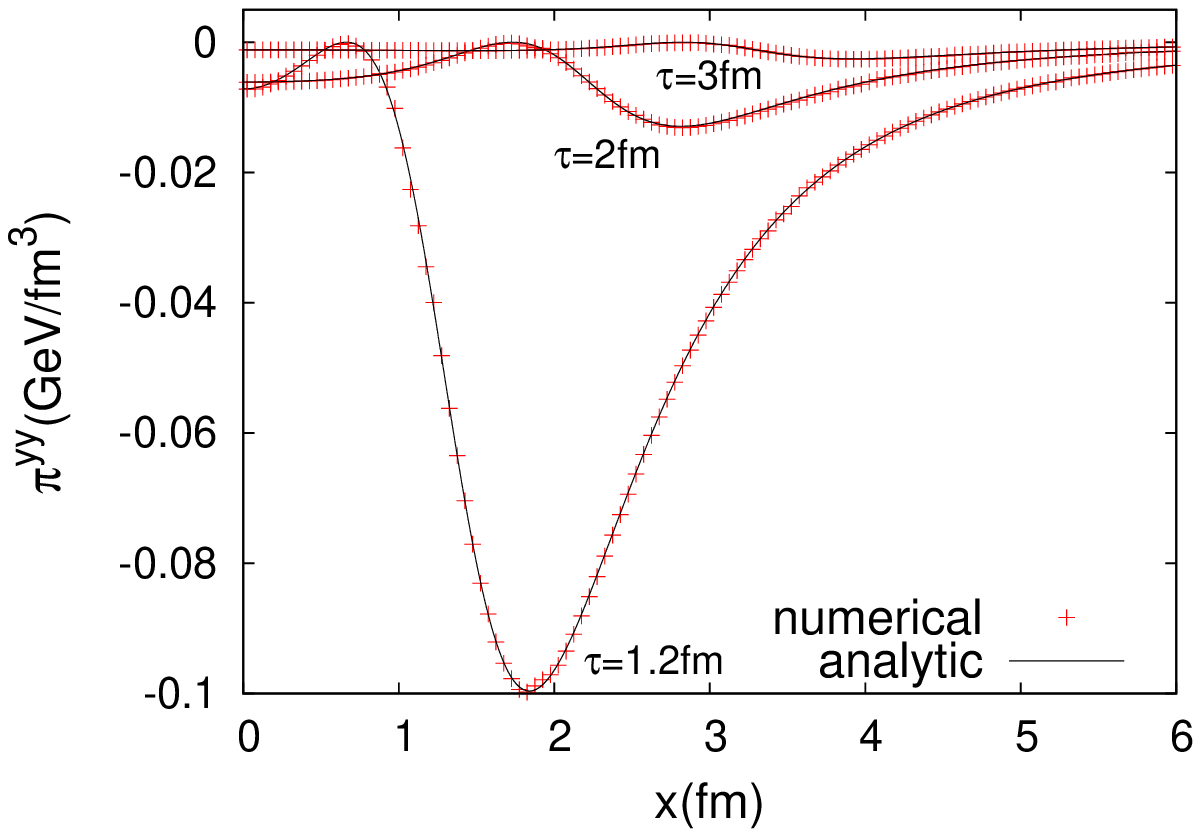}
  \end{minipage}

 \begin{minipage}{0.5\hsize}
  \centering
  \includegraphics[width=8cm]{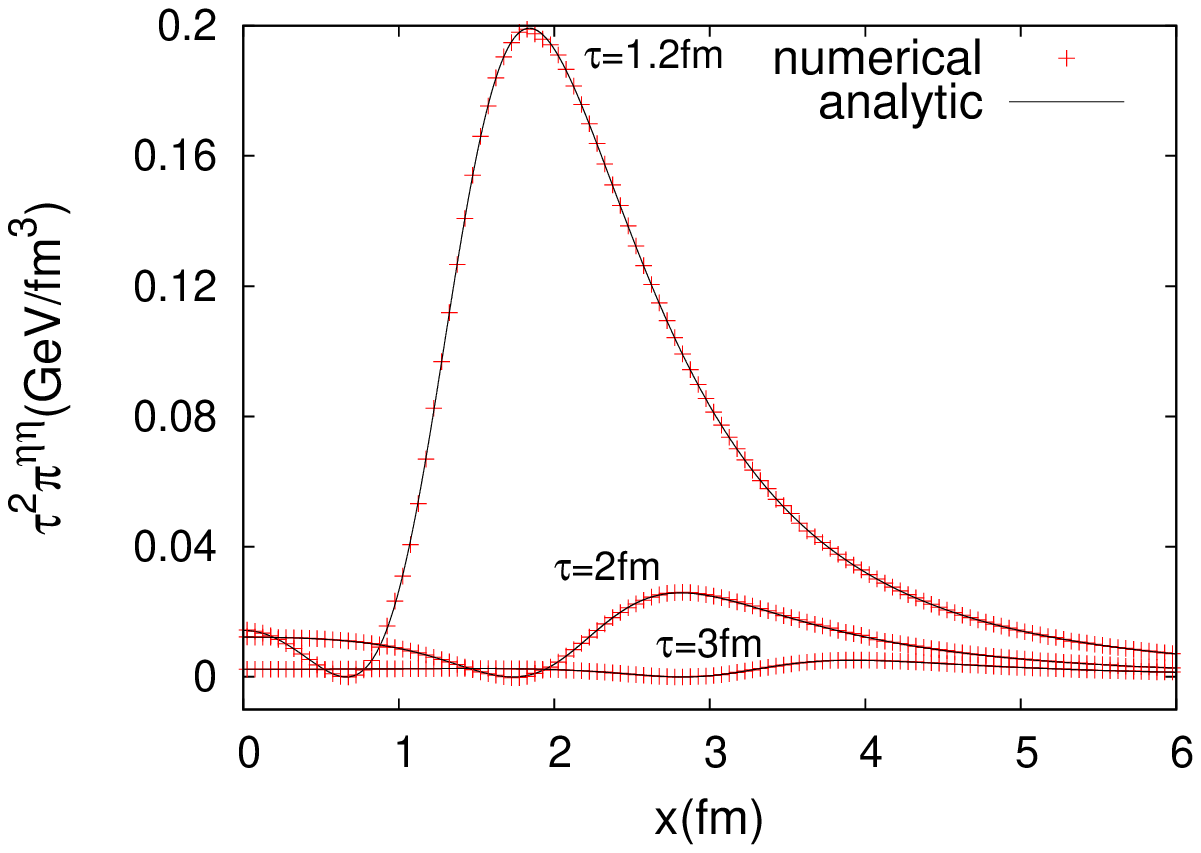}
  \end{minipage}
  \begin{minipage}{0.49\hsize}
  \centering
  \includegraphics[width=8cm]{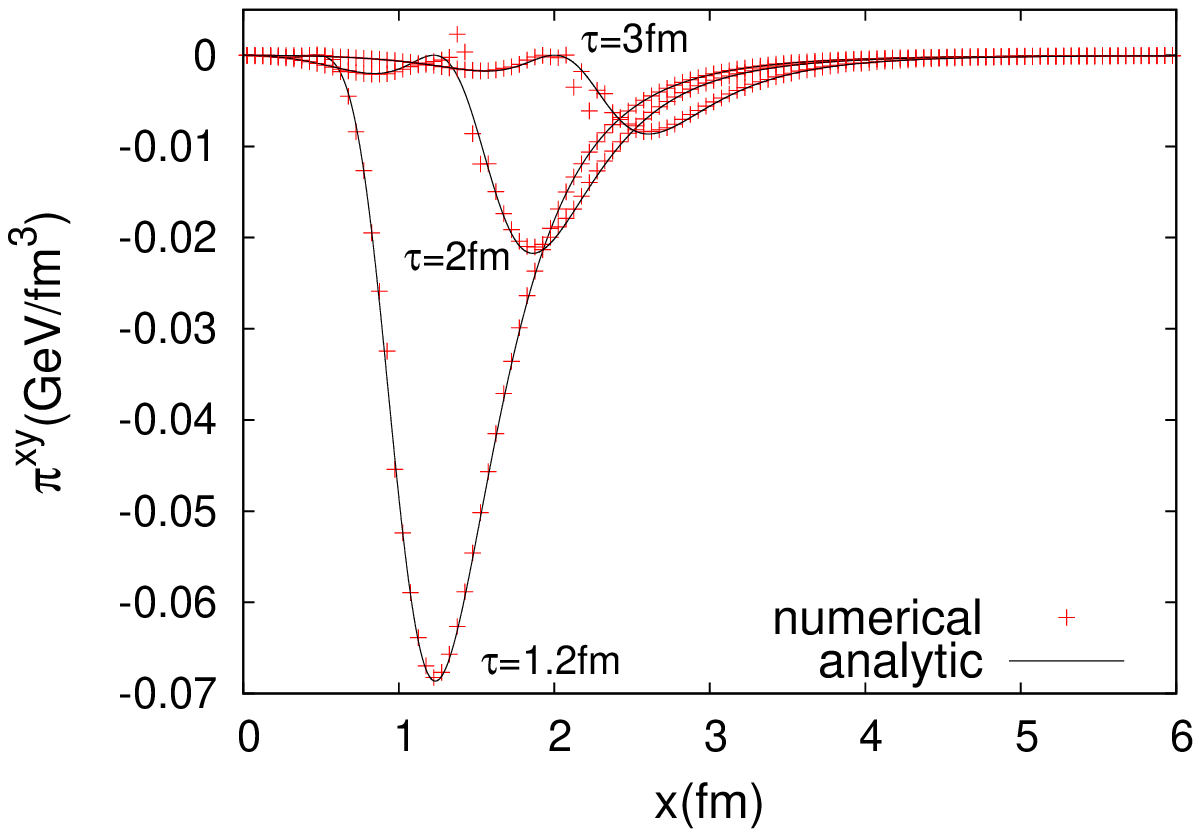}
  \end{minipage}
  \caption{
  Comparison between the solutions for shear tensors $\pi^{xx}$ (top left), $\pi^{yy}$ (top right), $\tau^2\pi^{\eta \eta}$ (bottom left) 
  and $\pi^{xy}$ (bottom right) from  the Gubser flow and our numerical calculation as a function of $x$.  
 The solid lines stand for the semi-analytic solutions and  the pluses stand for numerical results. 
\label{fig:Gubser-pi} }
\end{figure*}

In the numerical calculation, 
we set $\Pi_0=0$, $\zeta=1000$ MeV/fm$^2$, $\tau_0 = 1$ fm, 
and $\tau_\Pi=1$ fm. The space-grid size $\Delta\eta = 0.1$ and the time-step size $\Delta\tau = 0.1 \tau_0\Delta\eta$ are utilized. 
Figure \ref{fig:Bj-bv} shows the analytical and numerical results of the time evolution of the bulk pressure 
in the Bjorken flow. Our numerical calculation is consistent with the analytical solution. 


\subsection{Israel-Stewart theory in the Gubser flow regime} \label{Gubser-test}

\begin{figure*}[ht]
  \begin{minipage}{0.5\hsize}
  \centering
  \includegraphics[width=8cm]{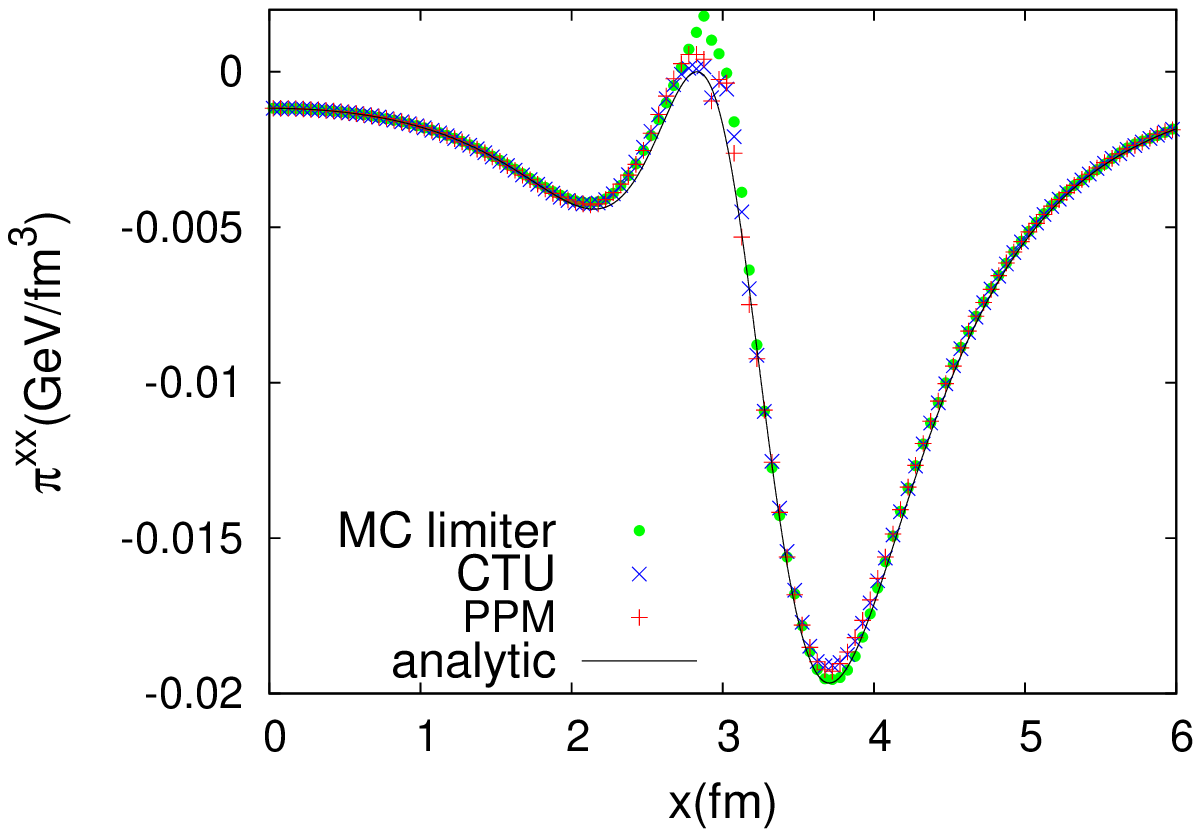}
  \end{minipage}
  \begin{minipage}{0.49\hsize}
  \centering
  \includegraphics[width=8cm]{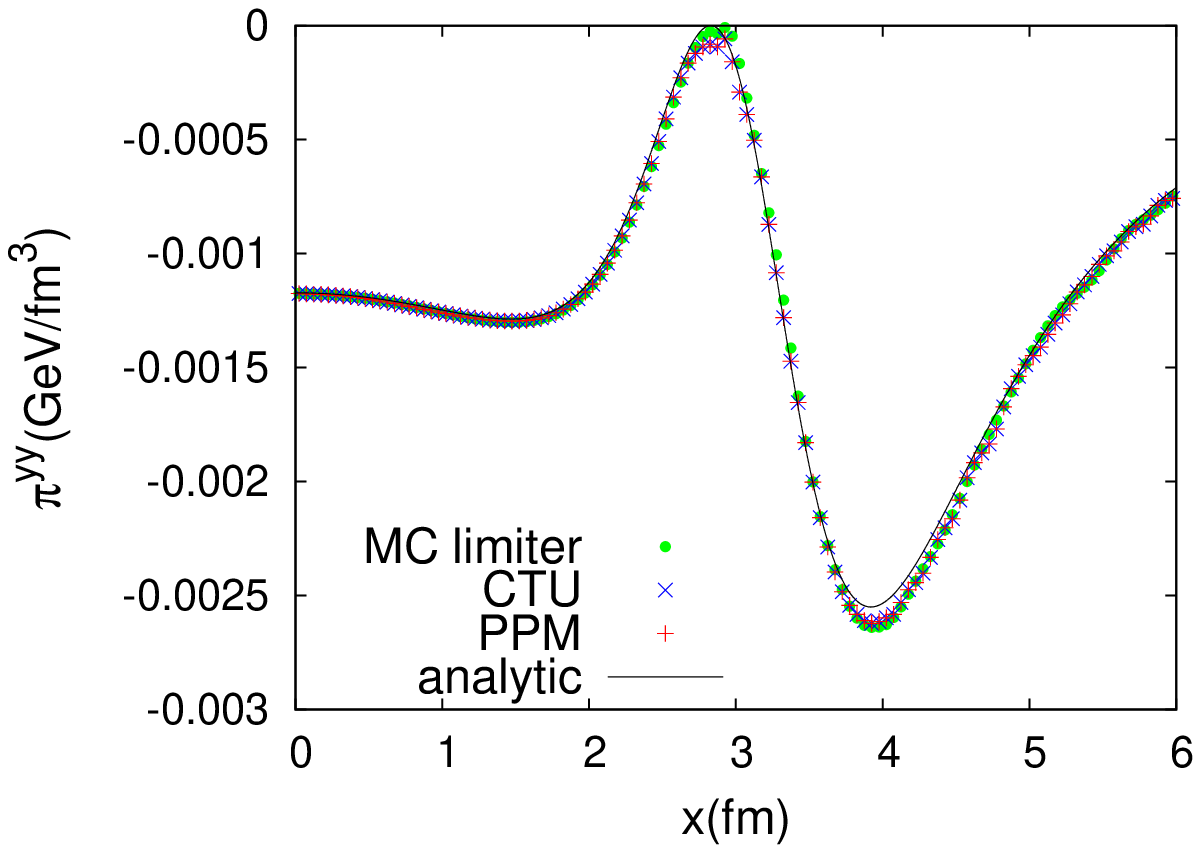}
  \end{minipage}

 \begin{minipage}{0.5\hsize}
  \centering
  \includegraphics[width=8cm]{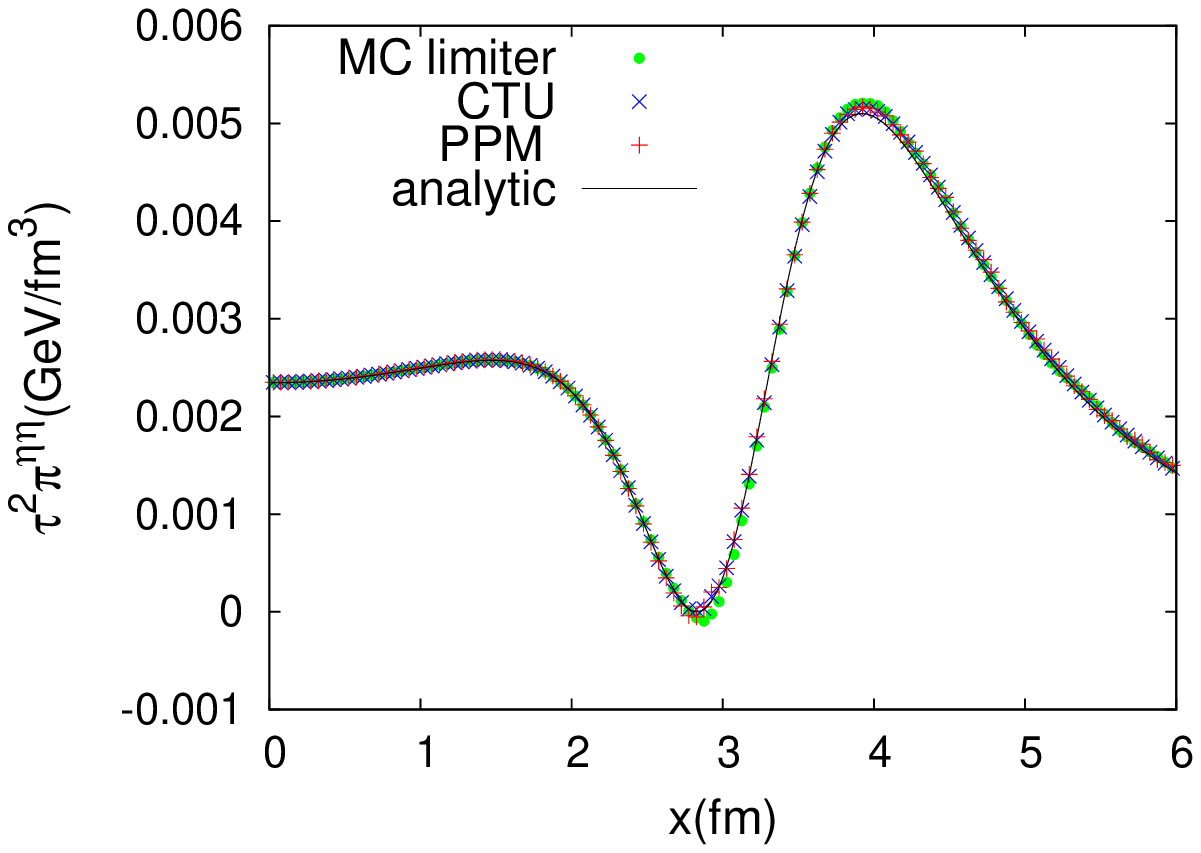}
  \end{minipage}
  \begin{minipage}{0.49\hsize}
  \centering
  \includegraphics[width=8cm]{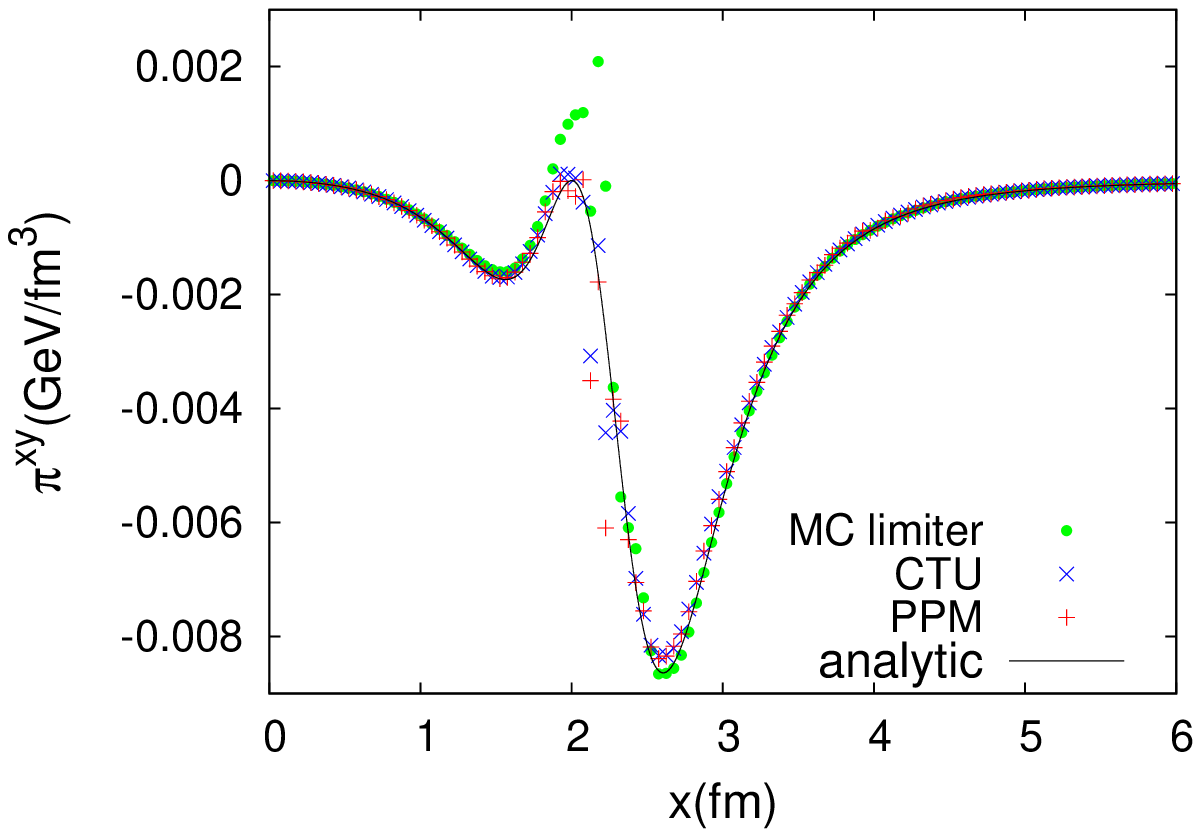}
  \end{minipage}
  \caption{Numerical results of shear tensors $\pi^{xx}$, $\pi^{yy}$, $\tau^2\pi^{\eta\eta}$  and $\pi^{xy}$ at $\tau=3$ fm 
  as a function of $x$, together with the semi-analytic solution (solid line). 
The solid circles, crosses and pluses denote the solutions obtained with the MC limiter, the CTU and the PPM method,  
respectively.  
\label{fig:Gubser-comp} }
\end{figure*}

Based on the symmetry arguments developed by Gubser \cite{Gubser2010,Gubser2011},  
a semi-analytic solution of 
the Israel-Stewart theory in the Gubser flow regime is obtained \cite{Marrochio2015}. 
The semi-analytic solution is a useful test problem for the code of relativistic 
viscous hydrodynamics which is developed for application to 
the high-energy heavy-ion collisions \cite{Marrochio2015, Noronha2014, Shen2014, Pang2014, ECHO2015}.
The velocity profile of the semi-analytic solution is the same as  that of 
the ideal Gubser flow,
\begin{equation} v^\bot =\frac{u^\bot}{u^\tau}=\frac{2q^2\tau x_\bot}{1+q^2\tau^2+q^2x_\bot^2} , \label{eq:gubserv}
\end{equation}
where $q$ is an arbitrary dimensional constant with unit of inverse length of the system size and set to $q=1$ 
in comparison with numerical computation. 
The solutions of the temperature and the shear tensors are derived by solving a set of two ordinary differential equations numerically \cite{Marrochio2015}. 
The second-order terms and the relaxation time in Eq.\:\eqref{eq:ISMilne2} are given by 
\begin{eqnarray} 
I^{\mu\nu}_\pi & = & \frac{4}{3} \pi^{\mu\nu} \theta,  \\ 
\tau_\eta & = & c \frac{\eta}{Ts}, 
\end{eqnarray}
where $c$ is a constant \cite{Marrochio2015}.

\par We carry out the numerical calculation with the finite shear viscosity $\eta/s = 0.2$. We set the relaxation time to $\tau_\eta=5\eta/(Ts)$. 
The numerical simulation starts at $\tau_0=1$ fm.
The time-step size and the space-grid size in numerical simulation are  set to 
$\Delta \tau= 0.1\Delta x$ and $(\Delta x, \Delta y, \Delta\eta )=(0.05 {\rm \hspace{1mm} fm}, 0.05 {\rm \hspace{1mm}fm}, 0.1)$, 
respectively. 

\par Figure \ref{fig:Gubser-T} shows the numerical results and the semi-analytic solutions of temperature and $x$ component of fluid velocity 
as a function of $x$ at $\tau=1.2$, 2 and 3 fm. 
The numerical results are consistent with the semi-analytic solutions. 
In our previous test calculation of the ideal Gubser flow the temperature and the fluid velocity follow the analytic solution 
until $\tau=7$ fm \cite{Okamoto2016}. 
On the other hand, in the finite viscosity calculation the difference between the numerical calculation and the semi-analytic 
solution appears after $\tau=4$ fm. 

In Fig.\:\ref{fig:Gubser-pi} the numerical results of the shear tensors $\pi^{xx}$, $\pi^{yy}$, $\pi^{\eta\eta}$  and $\pi^{xy}$  
at $\tau=1.2$, 2 and 3 fm 
are  presented together with the semi-analytic solutions. 
Here the profile of $\pi^{xy}$ is shown along a line $x=y$, since the value of $\pi^{xy}$ vanishes on 
the $x$ and $y$ axes.   
The shear tensors $\pi^{xx}$, $\pi^{yy}$ and $\pi^{\eta\eta}$ in our numerical calculations show good agreement 
with the semi-analytic solutions. 
However, in $\pi^{xy}$ the deviation from the semi-analytic solution starts to appear at $\tau=2$ fm and grows at later time. 

Since in the Israel-Stewart theory the second-order terms in $\pi^{\mu \nu}$ become small compared with the first-order terms, 
choice of numerical scheme for evaluation of the convection term in Eq.\:\eqref{eq:conv2} 
is important. 
For example, in Ref.\cite{Marrochio2015}, they show that 
adjustment of the flux limiter which controls possible artificial oscillation 
in a higher order discretization scheme is crucial for good agreement  with the semi-analytic solution. 
Here we employ the PPM for solving the convection part numerically, instead of the MC limiter. 
In the case of three-dimensional calculation we use the dimensional splitting method \cite{Okamoto2016}. 
We find that the Corner Transport Upwind (CTU) scheme \cite{Colella1990}  which is 
a three-dimensional unsplit method, realizes good agreement of the semi-analytic solution even with the MC limiter. 

We discuss the numerical scheme dependence on the shear tensors in 
solving the convection term in Eq.\:\eqref{eq:conv2}.  
We compare the three numerical schemes; a dimensional splitting method with 
the MC limiter, a dimensional splitting method with the PPM  
and the CTU method \cite{Colella1990} with the MC limiter for three-dimensional unsplit method. 
We shall explain the details of each scheme in \ref{app-interpolation} and \ref{app-convection}.
Figure \:\ref{fig:Gubser-comp} shows the semi-analytic solutions and numerical results of the shear tensors 
$\pi^{xx}$, $\pi^{yy}$, $\tau^2\pi^{\eta\eta}$  and $\pi^{xy}$ at $\tau=3$ fm. 
In $\pi^{xx}$, $\pi^{yy}$ and $\tau^2\pi^{\eta\eta}$, results of all numerical schemes 
are reasonably consistent with the semi-analytic solutions. 
In addition, differences among them are small. 
However, in $\pi^{xy}$ we can clearly see the scheme difference. 
In the solution obtained with the MC limiter, the large deviation from the semi-analytic solution at the peak around $x=2$ fm appears, 
whereas the PPM and the CTU methods keep the good agreement with the semi-analytic solution. 
The CTU method can achieve the high numerical accuracy  with the second-order accurate in space,
 but it needs the more computer 
memory than the dimensional splitting method with the PPM does. 
Therefore we employ the dimensional splitting method with the PPM for solving the convection term.

\section{Kelvin-Helmholtz instability in Bjorken expansion \label{Sec:KHI}}
We discuss the possible  development of  the KH instability 
in relativistic heavy-ion collisions.  
The KH instability is one of the hydrodynamic instabilities.
It occurs on the interface between two horizontal streams  which have different 
velocities \cite{Drazin1981}. 
If it takes place,  perturbations to the interface between fluids grow and result in vortex  formation.
In heavy-ion collisions, the color-flux tube structure in initial condition 
can be an origin of the KH instability; 
fluctuations in the longitudinal direction are amplified with the KH instability, 
however, vortex formation is not observed \cite{Csernai2012}. 
Recently  initial fluctuations and QGP expansion not only in the transverse 
direction but also in the longitudinal direction 
have attracted interest  \cite{CMS2015, STAR2017a, STAR2017b}. 
Using the new relativistic viscous hydrodynamics code  which has small numerical viscosity,  
we investigate the KH instability and vortex formation in heavy-ion collisions.

For simplicity, we focus on hydrodynamic expansion in the  $(x,\eta)$ plane. 
The heavy ion accelerated with high-energy still has about 1 fm width in the longitudinal direction  ($z$ direction) 
due to the uncertainty principle. 
In other words,  a thin disk  composed of large-$x$ partons is covered by a cloud of small-$x$ partons. 
As a result, in the high-energy heavy-ion collisions parton-parton interactions may take place in the area within around 
1 fm from $z=0$ fm.  
Then if we consider the color-flux tube structure in the initial condition, each color-flux tube may evolve from a different  
interaction point in $|z| < 1$ fm.

\begin{figure}[t!]
   \centering
  \includegraphics[width=7cm]{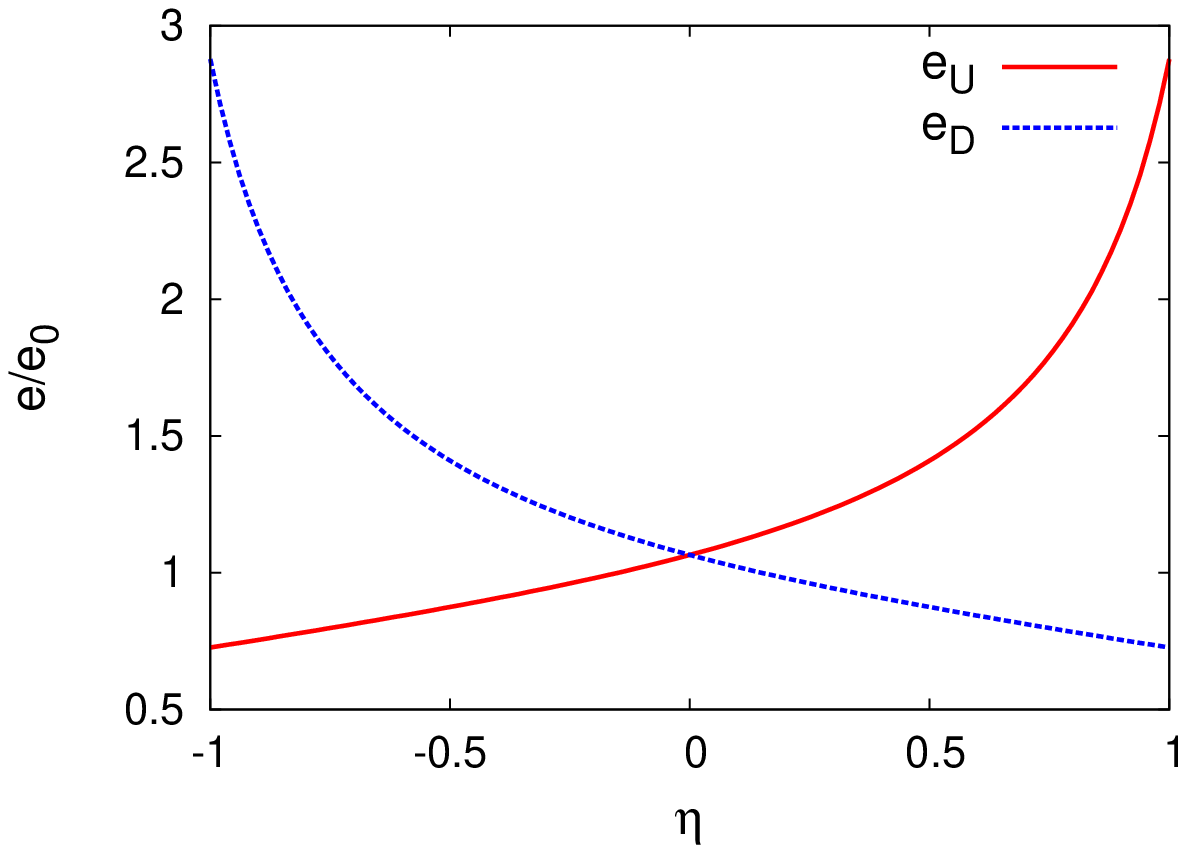}
  \centering
  \includegraphics[width=7cm]{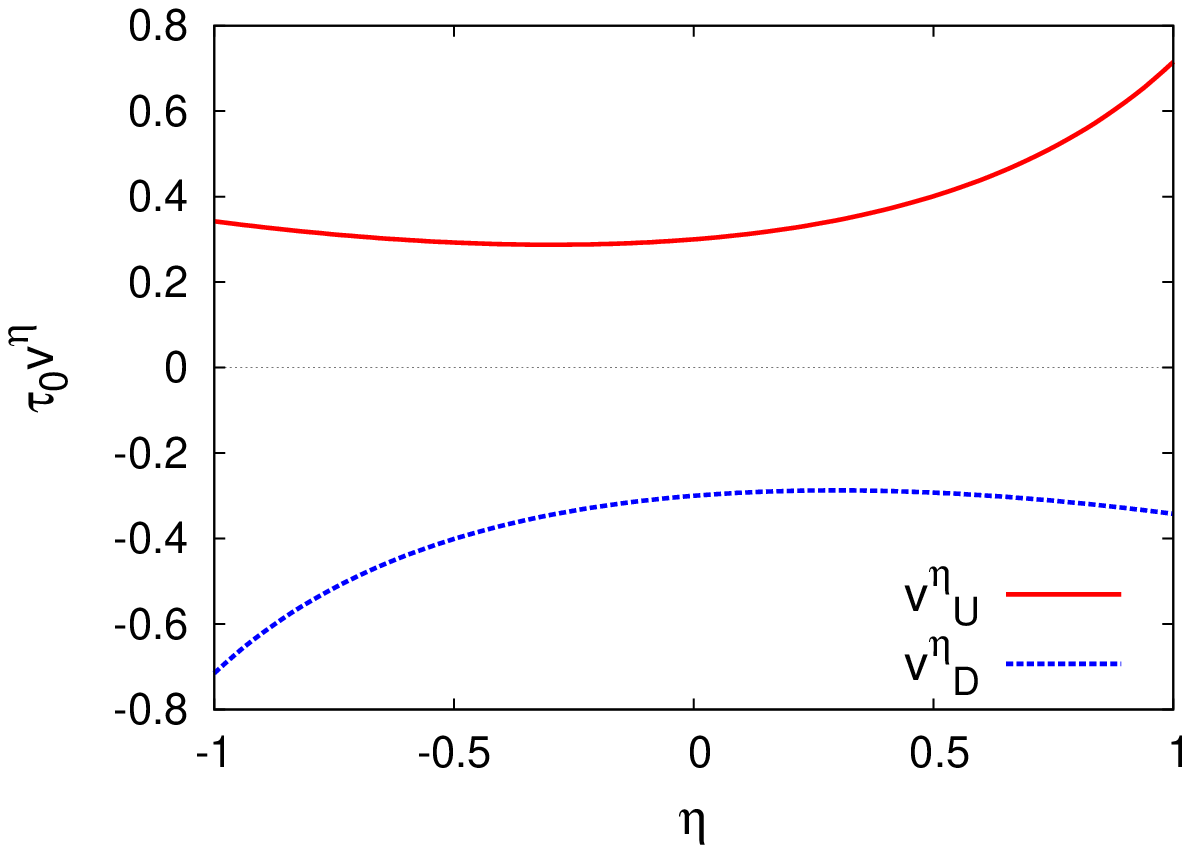} 
  \caption{The energy density (top panel) and the rapidity component of velocity (bottom panel) of the translated Bjorken flow. The results at $\tau= \tau_0=1$fm are shown. The red solid lines indicate the profile of Eqs.\:\eqref{eq:IC-KHeU} and \eqref{eq:IC-KHvU}. The blue dotted lines indicate the profile of Eqs.\:\eqref{eq:IC-KHeD} and \eqref{eq:IC-KHvD}.
\label{fig:shifted-Bjorken} }
\end{figure}

Suppose that two initial flow fluxes are located in $x>0$ and $x<0$ which represent two color-flux tubes starting 
to expand at $z=\Delta z$ and $z=-\Delta z$, respectively.  
Energy density and $\eta$ component of velocity of the flow flux are assumed to be described by Bjorken's scaling solution 
$e_B= e_0(\tau_0/\tau)^{4/3}$ and $v^\eta_B = 0$. 
Shifting the Bjorken scaling solution to $\pm \Delta z (=0.3 $ fm)  in the $z$ direction, 
we obtain energy density $e_{\rm U}$ ($e_{\rm D}$) and the $\eta$ component of velocity of the flow flux $v^\eta_{\rm U}$  ($v^\eta_{\rm D}$) 
in $x>0$  ($x<0$), 
\begin{align} 
e_{\rm U} (\tau,\eta) &= e_B(t, z+\Delta z) , \nonumber \\
& = e_0 \left( \frac{\tau_0}{ \sqrt{\tau^2 - 2\tau{\rm sinh}\eta \Delta z - \Delta z^2}}\right)^{4/3},  \label{eq:IC-KHeU}\\
 e_{\rm D}(\tau,\eta) &=e_B(t,z-\Delta z) \nonumber \\
&= e_0 \left( \frac{\tau_0}{ \sqrt{\tau^2 + 2\tau {\rm sinh}\eta\Delta z - \Delta z^2}}\right)^{4/3}, \label{eq:IC-KHeD}
\end{align} 
\begin{align}
v^\eta_{\rm U}(\tau,\eta) &= v^\eta_B(t, z+\Delta z) = \frac{\Delta z}{\tau^2} \frac{{\rm cosh}\eta}{ 1 - \frac{\Delta z}{\tau}{\rm sinh}\eta}  ,\label{eq:IC-KHvU}\\
 v^\eta_{\rm D} (\tau,\eta) & = v^\eta_B(t, z-\Delta z) = -\frac{\Delta z}{\tau^2} \frac{{\rm cosh}\eta}{ 1 + \frac{\Delta z}{\tau}{\rm sinh}\eta} ,  \label{eq:IC-KHvD} 
\end{align}
where $\tau_0$  and $e_0$ are the  initial time and the energy density, respectively.  
Figure \ref{fig:shifted-Bjorken} shows the energy densities $ e_{\rm U}$  and $e_{\rm D}$, and the
$\eta$ component of velocity of the flow flux $v^\eta_{\rm U}$ and $v^\eta_{\rm D}$ in $x>0$ and $x<0$. 
The energy density and the $\eta$ components of velocity of the flow flux are dependent on $\eta$ in $x>0$ and $x<0$, 
because of the translational transformation of Bjorken's scaling solution in the $z$ direction. 
Importantly, one can see that the shear flow is created between the two initial flow fluxes.

\begin{figure}[t!]
  \centering
  \includegraphics[width=8cm]{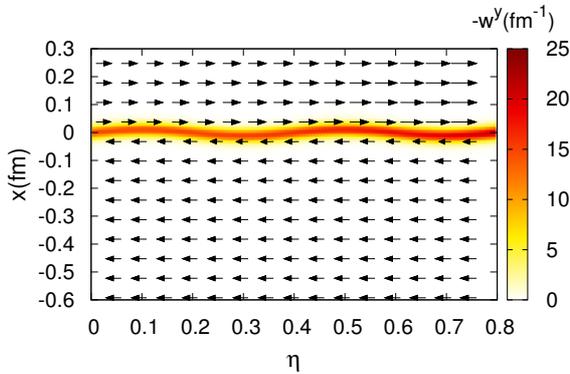}
  \caption{Initial condition for the shear flow with the Bjorken expansion.The color profile show the distribution of vorticity $-w^y$. The arrows indicate the three-fluid vector $(\tau v^\eta, v^x)$. 
\label{fig:KHI-IC} }
\end{figure}

Furthermore, we put the fluctuation $x_b= 0.01 {\rm sin}(2\pi\eta/\lambda)$ with a wavelength $\lambda$ 
along the boundary between the flow fluxes. 
Finally our initial energy density and flow velocity are written by 
\begin{align} e (\tau_0,x,\eta) &= \frac{e_{\rm U}(\tau_0,\eta) + e_{\rm D}(\tau_0,\eta) }{2}  \nonumber \\
& + \frac{e_{\rm U}(\tau_0,\eta) - e_{\rm D}(\tau_0,\eta)}{2}{\rm tanh}\left(\frac{x-x_b}{\Delta}\right) , \label{eq:initial-KH1} \\
 v^\eta(\tau_0,x, \eta) &= \frac{v^\eta_{\rm U}(\tau_0,\eta) + v^\eta_{\rm D}(\tau_0,\eta)}{2} \nonumber \\
 & + \frac{v^\eta_{\rm U}(\tau_0,\eta) - v^\eta_{\rm D}(\tau_0,\eta)}{2}{\rm tanh}\left(\frac{x-x_b}{\Delta}\right) , \label{eq:initial-KH2}
\end{align}
where the energy density and flow velocity around the boundary are connected 
from $e_U$ and $v^\eta_U$ to $e_D$ and $v^\eta_D$ smoothly with the parameter $\Delta$. 
Here, we set the wavelength of a fluctuation and the width of boundary between 
two fluid fluxes   to $\lambda=0.4$ and $\Delta = 0.02$ fm, respectively.  
Focusing the hot spots in a  fluctuating initial condition at the LHC,  we fix the initial 
energy density (temperature) to $e_0=741$ GeV/fm$^3$($T_0=800$ MeV). 
Figure \ref{fig:KHI-IC} shows the velocity field and profile of the vorticity $w^y$ of the initial condition 
Eqs.\:\eqref{eq:initial-KH1} and \eqref{eq:initial-KH2}. 
Here  the definition of the vorticity $w^y$, 
\begin{equation} w^y = \frac{1}{\tau}\left( \frac{\partial u^x}{\partial\eta} - \tau^2 \frac{\partial u^\eta}{\partial x}\right). 
\end{equation}
The arrows stand for the velocity field in $(\tau v^\eta, v^x)$. 

We start the numerical calculation on the grid $(\Delta x, \Delta\eta) =(0.005 {\rm\hspace{1mm} fm}, 0.00625) $ at $\tau_0=1$ fm  with 
time-step  size $\Delta\tau=0.2\Delta x$. 
We use the ideal gas equation of state $e=3p$.

\begin{figure*}[t!]
  \begin{minipage}{0.5\hsize}
  \centering
  \includegraphics[width=8cm]{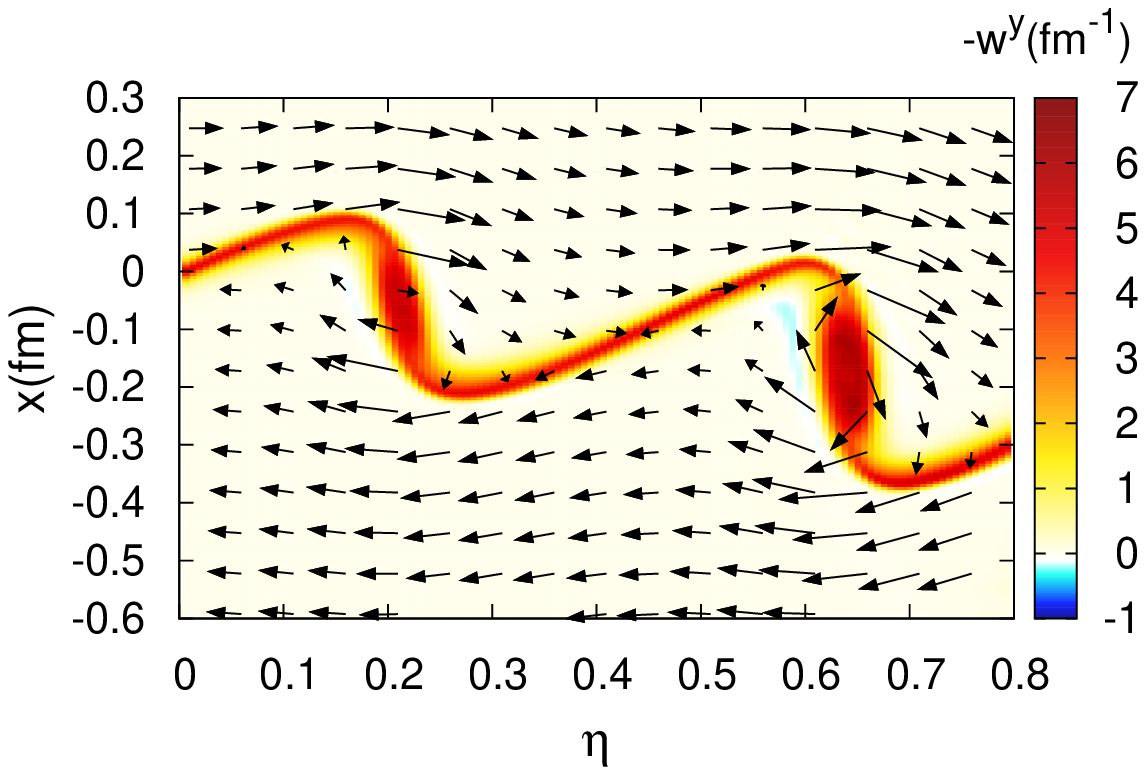}
  \end{minipage}
  \begin{minipage}{0.49\hsize}
  \centering
  \includegraphics[width=8cm]{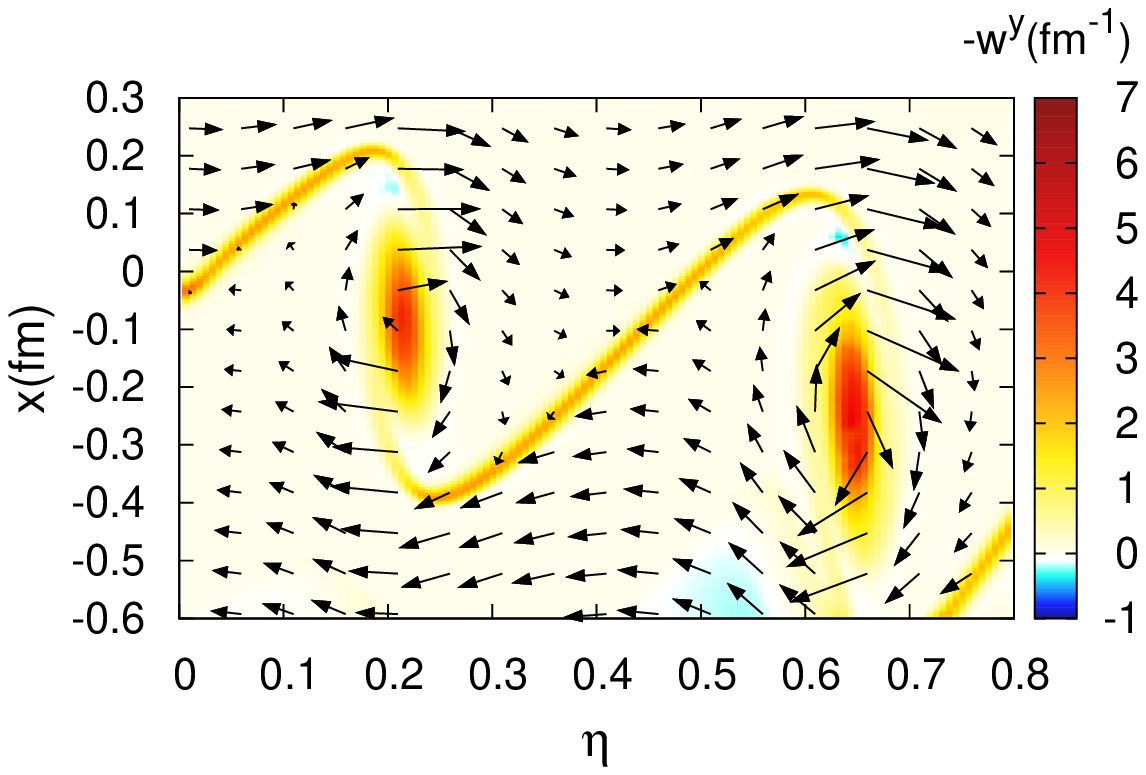}
  \end{minipage} 
  \caption{The evolution of KH instability in Bjorken expansion. The results of ideal fluid calculation at $\tau = 4$fm(left) and $7$fm(right) are shown. 
  The color profile indicates the distribution of vorticity $-w^y$. The arrows indicate the three-fluid vector $(\tau v^\eta, v^x)$. 
\label{fig:KHI-id} }
\end{figure*}
\begin{figure*}[t!]
  \begin{minipage}{0.5\hsize}
  \centering
  \includegraphics[width=8cm]{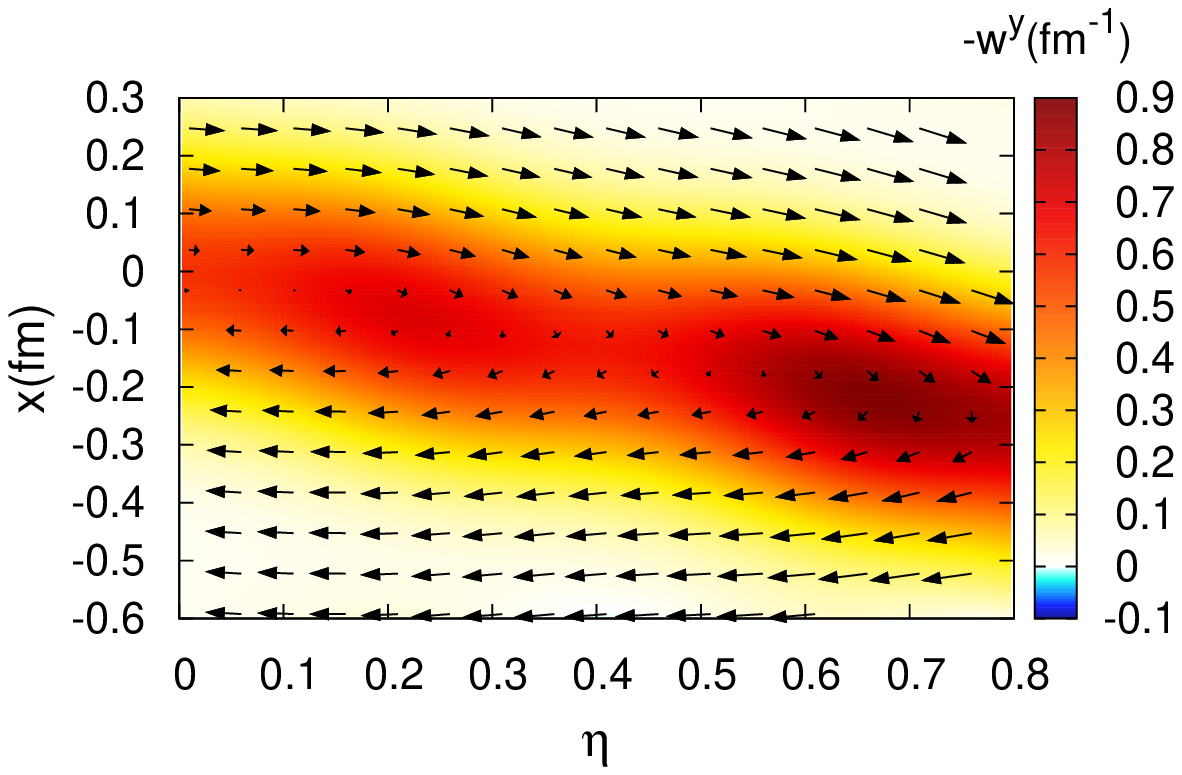}
  \end{minipage}
  \begin{minipage}{0.49\hsize}
  \centering
  \includegraphics[width=8cm]{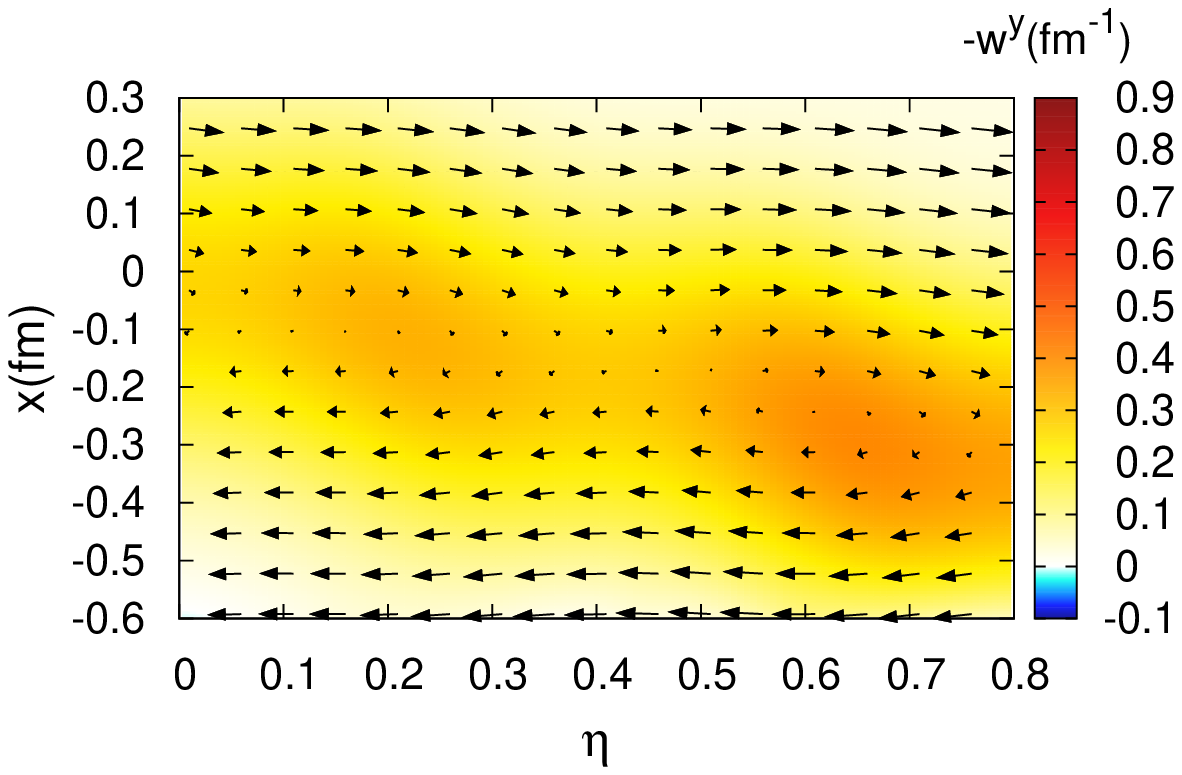}
  \end{minipage} 
  \caption{The evolution of KH instability in Bjorken expansion. The results of the viscous fluid calculation with $\eta/s=0.01$ at $\tau = 4$fm (left) and $7$fm (right) are shown. The color profile indicates the distribution of vorticity $-w^y$. The arrows indicate the three-fluid vector $(\tau v^\eta, v^x)$. 
\label{fig:KHI-vis} }
\end{figure*}

First we argue on the KH instability in the ideal fluid. 
We find a starting vortex formed around the boundary at $\tau \sim 3$ fm. 
In Fig.\:\ref{fig:KHI-id}  the velocity field and the profile of the vorticity $w^y$ at $\tau=$4 and 7 fm are shown. 
We observe that the boundary with the two vortexes tilt toward negative $x$. 
The initial conditions Eq.\:\eqref{eq:initial-KH1} and \eqref{eq:initial-KH2} and Fig.\:\ref{fig:shifted-Bjorken} suggest  
that $e_U$ is larger than  $e_D$ and $|v^z|$ in $x>0$ is larger than that in  $x>0$. 
The $e_U$ decreases more slowly than $e_D$  does due to the time dilation 
from larger $|v^z|$. 
The energy density and the flow differences between $x>0$ and $x<0$ cause the flow in the negative $x$ direction. 
The two vortices expand with time and their sizes grow 
because of existence of the Bjorken flow. 
As a result, the intensity of the vortices becomes small. 
The larger the difference of velocity in the shear flow is, the faster the growth of instability is. 
That is why the development of vortex at $\eta \sim 0.6$ is faster than that at $\eta \sim 0.2$. 
The fluctuation with a longer wavelength grows slower in the KH instability than that with a shorter 
wavelength does. 
If we set the wavelength $\lambda$ to $\lambda>0.5$ in the region $|\eta|<0.8$, the growth of fluctuation 
is too slow to form the vortex and the fluctuation is smeared with the Bjorken flow. 
However, at the forward rapidity, a  fluctuation with a long wavelength can survive to form a vortex.

Next we discuss the KH instability with finite viscosity. 
We employ the same values of the second-order term and the relaxation time in Eq.\:\eqref{eq:ISMilne2}  as 
those in Sec.\:\ref{Gubser-test}. 
The shear viscosity is set to $\eta/s=0.01$. 
Figure \ref{fig:KHI-vis}  shows the numerical results of KH instability at $\tau=4$ and 7 fm. 
In contrast to Fig.\:\ref{fig:KHI-id}, we cannot find the clear vortex but 
small and vague enhancement of vorticity around $\eta\sim 0.2$ and 0.6.  
Again we can see that the flow in the negative $x$ direction is produced. 
The width between two fluid fluxes expands and the fluctuation is washed away before it forms a  vortex 
because of the viscosity effect. 
The KH instability is not developed. 
In viscous fluid, a small size vortex compared with the Kolmogorov length scale is smeared by the viscosity and cannot exist. 
The fluctuation with the wavelength $\lambda= 0.4$  at $\tau_0=1$  fm may be smaller than the Kolmogorov length scale. 
In the mid rapidity, $|\eta|<0.8$, a fluctuation with longwave length disappears due to the Bjorken flow and 
a fluctuation with short wavelength is smeared by the viscosity. 
However, because at forward rapidity a fluctuation with long wavelength grows faster, 
there may be a chance that the KH instability occurs. 
Or if the longitudinal flow is smaller than Bjorken's flow, a fluctuation with long wavelength survives and can 
form a vortex. 
The existence of the KH instability depends on the viscosity and the flow distribution.

\section{Summary \label{Sec:sum}} 
In this paper we have developed the new relativistic viscous hydrodynamics code.  
In the code, we employed the Milne coordinates which are suitable for the initial strong longitudinal expansion at high-energy 
heavy-ion collisions. 
After the brief explanation of the relativistic viscous hydrodynamic equations, we showed the numerical algorithm
of the code which has the ideal part and the viscous part. 
For the ideal part we employed the Riemann solver with the two shock 
approximation which achieves stable calculation  even with the small numerical viscosity \cite{Akamatsu2014} and 
for the viscous part we utilized the PES method \cite{Inutsuka2011}. 
Because we found that the order of accurate in space in the convection part of the viscous part is 
important, we applied the PPM instead of the MC limiter to the convection part. 
Next we examined the validity of our code using two test calculations; the viscous Bjorken flow for the one-dimensional test and the Israel-Stewart theory in the Gubser flow regime for the three-dimensional test. 
In both tests, our numerical calculations showed good agreement with analytical solutions. 
Besides, we pointed out that in the Gubser flow the shear tensors are sensitive to numerical scheme. 
Finally, we discussed the possible vortex formation through the KH instability in high-energy heavy-ion 
collisions. We focused on the mid rapidity and started the numerical calculations with 
the simple initial conditions inspired by the color-flux tube structure of hot spots in fluctuating initial conditions. 
In the case of the ideal fluid we found the vortex formation after $\tau\sim 3$ fm,  
however, we did not observe the vortex formation in the viscous fluid even with very small viscosity. 
To obtain a more conclusive result for the vortex formation in high-energy heavy-ion collisions, 
we need to use the more realistic initial conditions. For example, the existence of shear flow is found in 
the initial condition based on the Color Glass Condensate \cite{Fries1, Fries2}. 
In addition, the effect of deviation from the Bjorken flow in a realistic initial condition is also important. 
Furthermore  we shall apply our new code to analyses of experimental data at RHIC and the LHC; 
correlation between flow harmonics \cite{ATLAS2015, ALICE2016},  event plane 
correlation \cite{ATLAS2014, CMS2015}, non-linearity of higher flow harmonics  \cite{Yan2015}  and 
three particle correlation \cite{STAR2017a, STAR2017b}. 
A comprehensive investigation of experimental data with the accurate numerical method 
of the relativistic viscous hydrodynamics gives us deep insight of QCD matter. 

\section*{Acknowledgments}
We would like to thank Berndt Mueller and Rainer Fries for valuable discussions.  
C.N. is  grateful for the hospitality of the members of Cyclotron Institute and Department of Physics and Astronomy in Texas A\&M. 
The work of C.N. is supported by the JSPS Grant-in-Aid for Scientific Research (S) No. 26220707 and 
US Department of Energy grant DE-FG02-05ER41367. 

\appendix

\def\thesection{Appendix \Alph{section}} 

\section{Interpolation procedures} \label{app-interpolation}
We give the explicit expressions for the interpolation procedure used in our relativistic viscous hydrodynamics code. 
We use the MC limiter for the second-order accurate in space and the PPM for the third-order accurate in space. 
We denote the center of the $i$th cell by $x_i$ and the boundary between the $i$th and the $i+1$th cell by $x_{i+1/2}$. 
We assume that we have the average value $a_i$ of the quantity $a(x)$ in the cell $(x_{i-1/2}, x_{i+1/2})$ 
where $a(x)$ stands for fluid variables and viscous tensors.
In the interpolation procedure, we evaluate the values of $a(x)$ at the right and left interfaces, 
$a_{R,i}={\rm lim}_{x\rightarrow x_{i+1/2} }a(x)$ and $a_{L,i}={\rm lim}_{x\rightarrow x_{x-1/2}}a(x)$ from 
the average value $a_i$. 

\subsection{MC limiter}
The second-order accuracy in space is achieved by the linear interpolation. 
In the second-order interpolation, we evaluate the interpolated values of $a(x)$ at right and left interfaces,  
\begin{align} a_{R,i} = a_i + \Delta a_i/2 ,\quad
a_{L,i} = a_i  -\Delta a_i/2.
\end{align}
In the MC limiter \cite{Van1979}, $\Delta a_i$ is given by
\begin{align} \Delta a_i = & {\rm min}( |a_{i+1} - a_{i-1}|/2 , 2|a_{i+1}-a_i|, 2|a_i - a_{i-1}| ) \nonumber \\ 
  & \times {\rm sign}(a_{i+1} - a_{i-1})  \quad {\rm if} \; (a_{i+1} - a_i)(a_i - a_{i-1})>0  , \nonumber \\
  = & 0 \quad {\rm otherwise} .  \label{eq:app-MC}
\end{align}

We define space averages of an interpolation function, $F_{i,R}(\sigma_i)$ and $F_{i,L}(\sigma_i)$,
\begin{align} F_{i,R} &= \frac{1}{\sigma_i\Delta x} \int^{x_{i+1/2}}_{x_{i+1/2} - \sigma_i\Delta x} a^I(x)dx , \\
F_{i,L} &= \frac{1}{\sigma_i\Delta x} \int^{x_{i-1/2}+\sigma_i\Delta x }_{x_{i-1/2}}a^I(x) dx, 
\end{align}
where $a^I(x)$ is an interpolation function of $a(x)$ and $\sigma_i = |u_i|\Delta t/\Delta x$. 
Here we use the sound velocity (the fluid velocity) for $u_i$ in the conservation equation (the convection equation). 
We utilize $F_{i,R}(\sigma_i)$ and $F_{i+1,L}(\sigma_{i+1})$ for the initial condition 
of the Riemann problem at the cell interface $x_{i+1/2}$ in the conservation equation. 
In the convection equation, $F_{i,R}(\sigma_i)$ or $F_{i+1,L}(\sigma_{i+1})$ corresponds to the numerical flux 
passing through the cell boundary $x_{i+1/2}$ (\ref{app-convection}). 
In the linear interpolation, $F_{i,R}(\sigma_i)$ and $F_{i,L}(\sigma_i)$ are expressed by
\begin{align} F_{i,R}(\sigma_i) =& a_{i,R} - \frac{\sigma_i\Delta x}{2}\frac{\Delta a_i}{\Delta x},  \label{eq:app-mc1}\\
F_{i,L} (\sigma_i)=& a_{i,L} + \frac{\sigma_i \Delta x}{2}\frac{\Delta a_i}{\Delta x}. \label{eq:app-mc2}
\end{align}

\subsection{Piecewise Parabolic Method (PPM)\cite{Colella1984, Marti1996,Colella2008}}
First, we calculate interpolated values of $a(x)$ at cell interfaces using forth-order interpolation.
\begin{align} a_{i+1/2} = \frac{7}{12}(a_i + a_{i+1}) - \frac{1}{12}(a_{i-1} + a_{i+2}) .
\end{align}
If the condition ${\rm min}( a_i, a_{i+1}) \leq a_{i+1/2} \leq {\rm max} (a_i, a_{i+1})$ is not satisfied, 
$a_{i+1/2}$ is limited as follows:
\begin{align} (D^2a)_{i+1/2} &= \frac{3}{\Delta x^2}(a_i - 2a_{i+1/2} + a_{i+1}) , \\
(D^2a)_{i+1/2,L} &= \frac{1}{\Delta x^2} (a_{i-1} - 2a_i + a_{i+1}) , \\
(D^2a)_{i+1/2,R} &= \frac{1}{\Delta x^2} (a_i - 2a_{i+1} + a_{i+2}) .
\end{align}
If the signs of $(D^2a)_{i+1/2}, (D^2 a)_{i+1/2,R}$ and $(D^2a)_{i+1/2 ,L}$ are all the same, 
\begin{align} (D^2a)_{i+1/2, {\rm lim} } =& {\rm min} \left( C| (D^2 a)_{i+1/2,L}| , C|(D^2a)_{i+1/2,R}|, \right. \nonumber \\
& \left. |(D^2a)_{i+1/2}| \right)  {\rm sign}( (D^2a)_{i+1/2}),
\end{align}
otherwise, $(D^2a)_{i+1/2,{\rm lim}}=0$. Then the modified values of $a_{i+1/2}$ read 
\begin{align} a_{i+1/2} \rightarrow \frac{1}{2} (a_i + a_{i+1}) - \frac{\Delta x^2}{3} (D^2a)_{i+1/2, {\rm lim}} ,
\end{align}
where $C>1$ is a constant. 
We set $C$ to $C=1.25$ \cite{Colella2008}. Then the interpolated values of $a(x)$ 
at right and left interfaces are initiated as $a_{L,i+1} = a_{R,i} = a_{i+1/2}$.

\par We perform the flattening algorithm near strong shocks to prevent numerical oscillations, 
\begin{align} a_{R,i} \rightarrow a_i f_i + a_{R,i}(1 - f_i) , \\
 a_{L,i} \rightarrow a_i f_i + a_{L,i} (1 - f_i) . 
\end{align}
The flattening parameter $f_i$ is fixed by $f_i = {\rm max}(\tilde{f}_i , \tilde{f}_{i+s_i})$, where $s_j = +1$ for $p_{i+1} - p_{i-1}>0$ and $s_j=-1$ for $p_{i+1} - p_{i-1}<0$, 
\begin{align} \tilde{f}_i = {\rm min} \left( 1, w_i {\rm max} \left( 0,\left(\frac{p_{i+1} - p_{i-1}}{p_{i+2} - p_{i-2} }-w^{(1)} \right)  w^{(2)} \right) \right) .
\end{align}
The constant $w_i$ is chosen by
\begin{align} w_i &= 1 \quad {\rm if}\; \frac{|p_{i+1} - p_{i-1}|} {{\rm min} (p_{i+1},p_{i-1})}  > \epsilon ,\; v_{i-1} > v_{i+1},  \nonumber \\
&= 0. \quad {\rm otherwise}
\end{align}
The parameters are set to $\epsilon = 1$, $w^{(1)}=0.52$, and $w^{(2)} = 10$ \cite{Marti1996}. The flattening algorithm is applied for conservation equations.

\par Furthermore, we modify the values of $a_{i,R}$ and $a_{i,L}$ to ensure the interpolated function remains monotonic. 
If $(a_{i,R} - a_i)(a_i - a_{i,L}) \leq 0$ or $(a_{i-1} - a_i)(a_i - a_{i+1}) \leq 0$, the $i$th cell contains a local extremum. 
The values of $a_{i, R}$ and $a_{i,L}$ are modified as follows:
\begin{align} (D^2 a)_i =& - \frac{2 a_{6,i}}{\Delta x^2},  \\
(D^2a)_{i,C} =& \frac{1}{\Delta x^2} (a_{i-1} - 2a_i + a_{i+1}) , \ \\
(D^2a)_{i,L} =& \frac{1}{\Delta x^2} (a_{i-2} - 2a_{i-1} + a_i) ,  \\
(D^2a)_{i,R} =& \frac{1}{\Delta x^2} (a_i - 2 a_{i+1} + a_{i+2}) ,  
\end{align}
where $a_{6,i}= 6a_i - 3(a_{i,L} + a_{i,R})$. If $(D^2a)_i$ and $(D^2a)_{i, \{L,C,R\}}$ have the same sign, 
\begin{align} (D^2a)_{i,{\rm lim}} =& {\rm min} (C|(D^2a)_{i,L}| , C|(D^2a)_{i,R}|, C|(D^2a)_{i,C}|,  \nonumber \\  
& |(D^2a)_i| ) {\rm sign} ((D^2a)_i) ,
\end{align}
otherwise, $(D^2a)_{i,{\rm lim}}=0$. Then we obtain 
\begin{align} a_{i, R} \rightarrow & a_i + (a_{i,R} - a_i) \frac{ (D^2a)_{i,{\rm lim}}} {(D^2a)_i }, \label{eq:extremum1} \\
a_{i, L} \rightarrow & a_i + (a_{i,L} - a_i) \frac{ (D^2a)_{i,{\rm lim}}} {(D^2a)_i }. \label{eq:extremum2}
\end{align}
If $(D^2a)_i=0$, we set the second term of Eqs.\:\eqref{eq:extremum1} and \eqref{eq:extremum2} to be zero. In the last limiter, the values of $a_{i,R}$ and $a_{i,L}$ are modified as
\begin{align} a_{i,R} & \rightarrow a_i - 2(a_{i,L} - a_i) \quad {\rm if} \; |a_{i,R} - a_i| \geq 2|a_{i,L} - a_i| , \\
a_{i,L} & \rightarrow a_i - 2(a_{i,R} - a_i)  \quad {\rm if}\; |a_{i,L} - a_i| \geq 2|a_{i,R} - a_i|.
\end{align}
The space averages of a parabolic interpolant are written 
\begin{align} F_{i,R} (\sigma_i) =& a_{i,R} - \frac{\sigma_i}{2}\left( a_{i,R}-a_{i,L} - \left(1-\frac{2}{3}\sigma_i\right) a_{6,i}\right) , \label{eq:app-ppm1} \\
F_{i,L}(\sigma_i)  =& a_{i,L} + \frac{\sigma_i}{2}\left( a_{i,R} - a_{i,L} + \left( 1 - \frac{2}{3}\sigma_i \right) a_{6,i}\right). \label{eq:app-ppm2}
\end{align}
Again, $F_{i,R}(\sigma_i)$ and $F_{i+1,L}(\sigma_{i+1})$ are used for the initial condition of the Riemann problem at the cell interface $x_{i+1/2}$ 
in the conservation equation. In the convection equation, 
$F_{i,R}(\sigma_i)$ or $F_{i+1,L}(\sigma_{i+1})$ corresponds to the numerical flux passing through the cell boundary $x_{i+1/2}$ (\ref{app-convection}).

\section{Numerical schemes for convection equations} \label{app-convection}

\subsection{High-resolution upwind method} \label{app-convection-1}
We consider the one-dimensional convection equation,
\begin{align}  \frac{\partial a(t,x)}{\partial t} + u(x) \frac{\partial a(t,x)}{\partial x} = 0. \label{eq:app-conv1}
\end{align}
In the high-resolution upwind method, we obtain the solution of the convection equation Eq.\eqref{eq:app-conv1}, 
\begin{align} a^{n+1}_i = a^n_i - \frac{u_i\Delta t}{\Delta x} \left( a^{n+1/2}_{i+1/2} - a^{n+1/2}_{i-1/2}\right), 
\end{align}
where $a^n_i$ is the value of $a(t,x)$ at $(t,x)=(t^n ,x_i)$, 
$a^{n+1}_i$ is the value of $a$ at next time step $t=t^{n+1} = t^n+\Delta t$. 
The numerical flux $a^{n+1/2}_{i+1/2}$ reads 
\begin{align} a^{n+1/2}_{i+1/2} =& F_{i, R}(\sigma_i)\quad {\rm if}\; u_i >0, \nonumber \\
=& F_{i+1, L}(\sigma_{i+1})\quad {\rm otherwise}.  \label{eq:app-upwind-flux}
\end{align}
We evaluate the $F_{i,R}(\sigma_i)$ and $F_{i,L}(\sigma_i)$, using the MC limiter (Eqs.\:\eqref{eq:app-mc1} and \eqref{eq:app-mc2}) 
or the PPM (Eqs.\eqref{eq:app-ppm1} and \eqref{eq:app-ppm2}). 

In the case of multidimensional problems, we employ the Strang splitting method \cite{Strang1968}. 
Using the operator $L^k_i$, which represents one-dimensional evolution in the $i$ direction during the time $k\Delta t$, 
we express two-dimensional expansion in the $(x,y)$ coordinates as 
\begin{equation} 
a^{n+1} = L^{1/2}_x L_y^1 L_x^{1/2} a^n.
\end{equation}
Similarly the three-dimensional expansion in $(x,y,z)$ coordinates  is written by 
\begin{align} 
a^{n+1} &= L^{1/6}_x L^{1/6}_y L_z^{1/3}L^{1/6}_y L_x^{1/3} L_z^{1/6} L_y^{1/3} L_x^{1/6} \nonumber\\
&\quad \times L_z^{1/3} L_x^{1/6} L_y^{1/3} L_z^{1/6} L_x^{1/6} a^n.
\end{align}

\subsection{Corner transport upwind (CTU) scheme  \cite{Colella1990}}
We consider two-dimensional convection equation, 
\begin{align} \frac{\partial a(t,x,y)}{\partial t} + u(x,y) \frac{\partial a(t,x,y)}{\partial x} + v(x,y)\frac{\partial a(t,x,y)}{\partial y} = 0.  \label{eq:app-conv}
\end{align}
In the CTU, the solution of the convection equation Eq.\eqref{eq:app-conv} reads  
\begin{align} a^{n+1}_{i,j} =& a^n_{i,j} - \frac{u_{i,j}\Delta t}{\Delta x} (a^{n+1/2}_{i+1/2,j} - a^{n+1/2}_{i - 1/2,j})  \nonumber \\
 & - \frac{v_{i,j}\Delta t}{\Delta y} (a^{n+1/2}_{i,j+1/2} - a^{n+1/2}_{i,j-1/2}) ,
\end{align}
where $a^n_{i,j}$ is the value of $a(t,x,y)$ at $t=t^n , x=x_i, y=y_j$, $a^{n+1}_{i,j}$ is the value of $a(t,x,y)$ at next time step $t=t^{n+1} = t^n+\Delta t$, the second and third terms stand for the numerical flux passing through the cell boundary. The numerical flux is given by
\begin{align} a^{n+1/2}_{i+1/2,j} 
=& a^n_{i,j} + \left(\frac{\Delta x}{2} - u_{i,j} \frac{\Delta t}{2}\right) \frac{\Delta^x a_{i,j}}{\Delta x} \nonumber \\
&- {\rm max}(v_{i,j},0) \frac{\Delta t}{2\Delta y} (a^n_{i,j} - a^n_{i, j-1})  \nonumber \\
&- {\rm min}(v_{i,j},0) \frac{\Delta t}{2\Delta y} (a^n_{i,j+1} - a^n_{i,j}) \quad {\rm if} \; u_{i,j} \geq 0, \nonumber \\
=& a^n_{i+1,j} - \left(\frac{\Delta x}{2} + u_{i+1,j} \frac{\Delta t}{2}\right) \frac{\Delta^xa_{i+1,j}}{\Delta x} \nonumber \\
&- {\rm max} (v_{i+1,j},0) \frac{\Delta t}{2\Delta y} (a^n_{i+1,j} - a^n_{i+1, j-1}) \nonumber \\ 
&- {\rm min} (v_{i+1,j},0) \frac{\Delta t}{2\Delta y} (a^n_{i+1,j+1} - a^n_{i+1,j})
\quad {\rm if} \; u_{i,j} < 0,\\
a^{n+1/2}_{i,j+1/2} =& 
a^n_{i,j} + \left(\frac{\Delta y}{2} - v_{i,j} \frac{\Delta t}{2}\right) \frac{\Delta^y a_{i,j}}{\Delta y} \nonumber \\
&- {\rm max}(u_{i,j},0) \frac{\Delta t}{2\Delta x} (a^n_{i,j} - a^n_{i-1, j})  \nonumber \\
&- {\rm min}(u_{i,j},0) \frac{\Delta t}{2\Delta x} (a^n_{i+1,j} - a^n_{i,j}) \quad {\rm if} \; v_{i,j} \geq 0, \nonumber \\
=&
 a^n_{i,j+1} - \left(\frac{\Delta y}{2} + v_{i,j+1} \frac{\Delta t}{2}\right) \frac{\Delta^y a_{i,j+1}}{\Delta y} \nonumber \\
& - {\rm max} (u_{i,j+1}, 0) \frac{\Delta t}{2\Delta y} (a^n_{i,j+1} - a^n_{i-1, j+1}) \nonumber \\ 
& - {\rm min} (u_{i,j+1}, 0) \frac{\Delta t}{2\Delta y} (a^n_{i+1,j+1} - a^n_{i,j+1})
\quad {\rm if} \; v_{i,j} < 0. 
\end{align}
Here we evaluate the variation of $a(t,x,y)$ in the $x$ ($\Delta^x a_{i,j}$)  and  $y$ direction ($\Delta^ya_{i,j}$) 
using the MC limiter \eqref{eq:app-MC}.


\end{document}